# UNIVERSAL CONCEPT OF COMPLEXITY
# BY THE DYNAMIC REDUNDANCE PARADIGM:
## Causal Randomness, Complete Wave Mechanics,
## and the Ultimate Unification of Knowledge

**Andrei P. Kirilyuk**[*]

Institute of Metal Physics, Kiev

**Abstract.** This is a brief, non-technical presentation of the main results of a book with the same title in which a new, rigorously defined concept of dynamic complexity is introduced and it is shown that it gives the complete and absolutely universal description of both representative particular cases of complex dynamics and arbitrary dynamical system behaviour. This crucial extension with respect to the existing concepts is achieved due to a new, universal method of analysis of arbitrary dynamic equations, avoiding the usual limitations of the essentially perturbative, one-dimensional approach of the canonical, linear (unitary) science (including any superficially defined, integrable 'nonlinearities'). The nonperturbative analysis shows that an equation describing any real behaviour with more than one effective dimension possesses many solutions, each of them being complete in the usual sense and approximately equivalent to some ordinary, 'exact' solution of the unitary science. Therefore these elementary complete solutions, called realisations, are incompatible among them and, being equivalent and thus equally probable, permanently and spontaneously replace one another, in the form of the corresponding dynamic regimes. This discovery, referred to as the dynamic redundance paradigm, provides a qualitatively new understanding of the notion of existence itself and universally explains all the known patterns of dynamic behaviour within the ensuing single concept. It provides, in particular, the causal, dynamically based and consistent definition of randomness and probability (or fundamental dynamic uncertainty) which appear in a closed, independent and irreducible form, and therefore are omnipresent. Being applied at the level of micro-objects, this concept gives the physically realistic, complete extension of quantum mechanics, equivalent to the unreduced version of de Broglie's double solution, amplified with the inherent dynamical chaos and intrinsically unified by its very origin with the extended, causal interpretations of 'special and general relativity', 'quantum gravity', 'field theory', 'particle physics', and cosmology. Among the numerous emerging particular results one can mention the realistic and universal interpretation of inertial mass-energy, inseparable from its gravitational interpretation, and physically real space and time. One can consistently describe the exact physical, realistic nature of an elementary particle and its properties, in full agreement between the known empirical manifestations and (extended) mathematical presentation. The ensuing intrinsic unification of knowledge is further extended to higher 'levels of complexity' including any dynamical system and their full ensemble which forms the omnipresent, fractally structured, and intrinsically self-developing hierarchy of dynamic complexity of the world, adequately described by the resulting 'science of complexity'. This is expressed, in particular, by the obtained unified Lagrange-Hamilton-Schroedinger formalism with the accompanying 'multivalued' and probabilistic interpretation that gives, in the corresponding limit, the extended, complex-dynamical version of any equation describing a real process. This single universal equation of the science of complexity is based on the equally unique and universal principle, the complexity conservation law (or the universal symmetry of complexity) which is confirmed by all known observations. It combines and considerably extends such canonical principles as ordinary conservation laws (energy, momentum, electric charge, etc.), the second law of thermodynamics, the 'principle of relativity', and various versions of the 'variational principle' (least action, minimum potential energy, etc.). Practically each basic concept of the canonical science acquires, within the science of complexity, its crucial, causally dynamic extension and realistic completion, and among them randomness (and the related probability, event, uncertainty, unpredictability, instability), dynamical chaos, structure and its 'spontaneous' formation (creation), irreversibility and discreteness, fractality, time and space, nonlinearity and interaction (entanglement), (non)separability, (non)integrability, general solution (completeness). Finally, the intrinsic unification of all the diverse fields of the canonical knowledge, in their extended versions, is outlined and demonstrated for various real problems from different fields including the humanities and theology at the highest perceived levels of the universal hierarchy of complexity. The universal science of complexity provides a qualitatively new type of knowledge, ultimately unified and inseparable from the reality it reflects. The advent of the extended thinking of the universal science of complexity confirms the end of the canonical, unitary type of knowledge and opens the renaissance of the ultimately complete and universal understanding initiated by René Descartes but then mechanistically falsified and unjustly discarded by the dominating linear approach of the scholar science. Many practically important conclusions are deduced within the new understanding and others are outlined.

---

[*] Address for correspondence: Post Box 115, Kiev - 30, Ukraine 252030.
 E-mail address: kiril@metfiz.freenet.kiev.ua.

# Prologue: *The End*

Complexity, nonlinearity, chaos, self-organisation, criticality... The flood of stylish words, pretentious publications, and advanced study centres originates from a new hope to create the unified science of complexity and explain at last everything within a single approach reproducing the intrinsic unity of Nature. However, the proposed concepts fail, one after another, and the predicted universality definitely escapes the most sophisticated developments of the mechanistic science including the highest supercomputer powers (Horgan 1995).

But the Universal Truth has seemed to be so close, with its vague contours already emerging from the disappearing mist of ignorance, and it remains the more attractive the more it resists to the massive attacks of the heavily armed formalism. Something qualitatively new is needed to see it, something fundamentally different, universal and therefore probably not so intricate in its form. One does not need a sophisticated key to open a tricky lock, the entrance is free, one just needs another vision to see it. The Truth reveals itself only to those who already have the germ of it inside their minds.

In the meanwhile, the canonical, or linear, science has entered the phase of absolute and helpless stagnation which is only emphasised by the growing success of certain its practical applications. The life of an idea does not stop with its discovery and scientific elaboration, it is simply transformed from a fundamental revelation into a practically useful instrument. There are now only two contrasting types of observations, separated by an abyss: those considered to be perfectly understood and successfully used and those which cannot be understood at all despite the truly gigantic efforts applied and independent of their possible practical use. The rupture between the two is so unreasonably insuperable that it seems sometimes to be irrational.

This is the ***End of Science***, the complete saturation of the canonical, mechanistic, unitary (linear) science that we knew until now, since it represents practically all the existing knowledge that is ordered enough to be classified as science (cf. Horgan (1996)).

The sophistication of modern electronics exceeds systematically any imagination, but the operation of elementary natural machines, like living cells, remains far beyond the possibilities of science. The simplest viruses are merely big organic molecules, but they can already do what none of artificial machines can - reproduce themselves by their own functioning - and all the power of modern science fails to control their single species.

Still much more simple, physical systems also easily escape the proclaimed omnipotence of the scholar science: although many people have become prosperous due to the astronomical investments into the field of high-temperature superconductivity, its mechanism remains unclear, after many years of very intensive investigation. The novelties in this and many other fields of physics appear as a result of a basically empirical, intuitive search resembling the more and more a modern version of the glorious alchemy. And where are the promised and generously sponsored inexhaustible sources of energy, like the controlled nuclear fusion, or really intelligent, 'thinking' computers? There is a whole list of the announced scientific miracles that have evidently crashed upon the same barrier of cognition, clearly seen now, after the initial period of apparently 'promising' development.

Finally, at the very basement of the universe we find the same impenetrable barrier, resisting to all human forces. The mysteries of quantum mechanics remain as unsolvable as they were at its origin, over 70 years ago, but now they are *provocatively* puzzling. The announced 'Great Unifications' of the elements of Being and Theories of Everything are transformed into a gibberish of artificial, abstract symbols where *everything* is indeed possible, as they are *irreducibly* separated from reality. The famous 'unreasonable effectiveness of mathematics in the physical sciences' has become a simply *unreasonable* sophistication, effectiveness left apart. It is clearly demonstrated by the easily performed comparison between the enormous volumes and generous support of directions like 'mathematical physics' and the total absence of a consistent, physically sound solution to any nontrivial real problem they are supposed to induce.

Clear functional signs of profound corruption are accumulating behind the technocratically maintained facade of the official science: omnipresent mediocrity and the veritable scientific parasitism actively suppressing the remnants of the genuine creativity; organised groups of dealers fighting for their personal promotion and using all kind of heavily formalised, 'scientifically looking' imitations of truth; proliferating unlimited 'blurring' of the indispensable and formerly firm ethical norms, - such are typical tendencies of degradation, consuming the whole institutions and fields of knowledge and emphasised by many serious and variously 'oriented' professionals in science and beyond (e. g. Maddox (1995a,b), Berezin (1996), Braben (1996), Farge (1996), Gross, Levitt, and Lewis (1996), Sangalli (1996), Ziman (1996), Bricmont (1997), Lawrence and Locke (1997), Postel-Vinay (1997), de Rosnay (1997), Sokal and Bricmont (1997), Wenneras and Wold (1997), Balter (1998)).

Probably the most meaningful and impressive sign of death of the canonical science, somehow summarising all the particular features of its stagnation, is the clearly seen loss of interest in it from various people, represented both by narrow specialists and the general public, by society in the whole and individual enthusiasts. It is as useless to try to reanimate artificially the interest in the dead canonical science as that science itself. The impasse of that scale cannot be avoided just by mechanically pumping milliards into the straightforward attack and formal publicity campaigns, while this does produce enough harm by attracting too much force from the search for a qualitatively new approach that can alone lead to the issue. For an issue always exists, but this time it can be found only within quite a new type of thinking whose universality should comprise, by definition, at least the whole diversity of the existing knowledge.

The End of Science is just a particular manifestation of the overall saturation in the civilisation development. Indeed, who can seriously believe that such fundamental conflict in the knowledge acquisition is closed within itself? After all, everything in human activity is the search for a new knowledge, irrespective of the accepted definition of science.

The economical, social, and cultural development of the world, as it is directly represented by the state of the 'developed' countries, has attained the same stage of fundamental local exhaustion as the scientific progress. It is characterised by the obvious global stagnation, the absence of a well-defined general direction in the dynamics of a system, this indispensable sign of its progressive development. Moreover, the decadent, descending motion, inevitably replacing the absent progress, is clearly discernible in the modern world dynamics.

The resulting **End of History** (cf. Fukuyama (1992)) manifests itself as the apparent absence of 'historical', qualitatively big events and not less obvious impossibility of their initiation within the current *mode d'existence*. The World has become 'lazy' and indifferent, it likes the show, the spectacle, but not the participation in the real events, whatever their contents is. The world of 'general consumption' cannot propose any genuine event in principle, it flickers around 'comfort' attained by 'arrangements' within the antagonistic 'competition of influences' and opposed to the true creation and event emergence. Proclaiming the 'decisive' refusal from 'bad' events, it rejects in reality any event, and thus any real progress that can only be based on the irreducible, 'big' creation, in science as well as in life in general. The genuine creation and the resulting richness of content of the 'true things' are replaced with the mechanistic change of external 'nice looking' forms which hide the dark emptiness of low, selfish 'interests'. Mediocre, treacherous stereotypes dominate the individual and collective behaviours at all scales: "all is vanity and vexation of spirit". A 'democratic' hierarchy of rigid mechanistic subordination actually replaces the formally announced 'general' freedom and gives rise to society of 'one-dimensional men' enslaved by the absolutely dominating, technocratic "beast" of the Unitary System, which shows once more that the real power cannot belong to everybody and everything.

The **End of Morals** and the related **End of Belief**, either religious, or ideological, or involving any other general conviction, stem from the same root. The deceptive replacement of moral values with the abstract, absolute liberty-equality detached from any preferred choice, and of a conscious belief with a crude fanaticism, only emphasize the reality of the End. In any living, developing system the necessary initial liberty is inseparable from its auto-limitation by a subsequent properly made choice; liberty of a living world means the freedom to choose its proper limitations leading to progressive changes that should indeed be realised in order to pass to a higher level of freedom (complexity), with the new really performed choices, and so on. It was always happening in this way in the most prosperous societies, provided they were not at their End. From the other hand, any most fervent belief can realise its goals only when it is creative, and creativity necessary implies much freedom.

The **End of Art** is the result of the related depletion of the 'free space' in the higher spheres of human imagination, which is especially similar to the End of Science. Everything that could be created, said and figured out *at the current level* has already been realised, only the details can be varied infinitely.

All the related aspects of the End appear in the form of profound, hopeless indifference and corruption of values, dominating unlimited mediocrity, cynicism and ultimate simplification, irrespective of the artificially maintained quality of 'facades'. The world is transformed into one big and mercenary 'show-business' in which everything turns out to be unreal: the events and heroes, stars and leaders, values and feelings are all as if played by mediocre actors under the guidance of a mediocre director, and the more pretentious are the forms, covers and envelopes, the more obvious is the emptiness they hide. Inside all the plays of words and instinctive self-protections, the dwellers of the End know well that they have nothing more to propose and seriously believe in, and many even guess that they can never return to the previous unconditional faith and 'expected' type of truth. The end of the Unitary System of life, power and thinking becomes perfectly complete.

***The End*** cannot be simply cured or destroyed, it is a state of destruction and a remedy itself. The local End, the exhaustion of the current level of development, can be surpassed only by transition to a new, qualitatively different stage of development; or else it will inevitably become the definite, absolute End of a completely exhausted civilisation, followed by the unavoidable demolition, and the universal development will restart again from lower levels, here or somewhere else.

The current End of Science can give rise to a new, much more extended type of truly scientific, conscious and unified knowledge, provided those who are involved are ready to welcome and develop the change that will necessarily be both progressive and dramatic in its contents and character. In fact, this new type of knowledge can only be the Universal Science of Complexity representing the Omniscience, a really omnipotent Theory of Everything, equivalent to the universal complete *understanding* which naturally combines theoretical consistency with practical mastership and unifies in principle *all* human knowledge, and not only the fundamentals of physics or conventionally interpreted sciences of 'exact' type. Everything that is less than this Ultimate Unification of Knowledge will not stop the End, since it will necessarily belong to the existing, conventional knowledge formed from portions of quasi-continuous logical chains with incomprehensible ruptures among them, so strikingly opposed to the unified harmony of Natural Being. All the basic potentialities of that 'half-conscious' state of scientific knowledge have already been explored, and the result is the End.

Due to its universality, the genuine Science of Complexity is also a way to the next big stage of the general civilisation development passing by the unique issue from the decadence of the local End that dangerously approaches now to its global, catastrophic version. This new type of science cannot be only a 'classified stock of knowledge', it is a qualitatively higher *way of thinking* and the ensuing superior way of living for *everybody*. One cannot avoid this mentally based *Revolution of Complexity* which relates inseparably all the 'degrees of freedom' of the civilisation and is the Beginning of a new level of consciousness. The latter has been anticipated by many, in various forms, but now it should be specified and implemented in the real life, in a well understood and unified form which is actually unique as such.

This work introduces the Universal Science of Complexity. Contrary to other existing attempts, remaining fundamentally within the same general approach of the canonical science and therefore condemned to failure, we start with a new paradigm, representing that 'something absolutely new' which seems to be indispensable for the successful unification. This is the paradigm of the *dynamic redundance*, or fundamental multivaluedness of dynamical functions (FMDF), naturally emerging in the formal description of any dynamic behaviour as the universal extension of the essential single-valuedness, artificially imposed by the canonical, mechanistic way of thinking and directly related to its dynamically linear, effectively one-dimensional analysis.

The discovered dynamic multivaluedness leads to the universal concept of *dynamic complexity*. Not only does it provide the complete description of complex behaviour of any real system, but also reveals the global structure of the world as the integrated hierarchical tree of levels of complex dynamics characterised, instead of ruptures, by the intrinsic *dynamical randomness* which leads to permanent appearance and growth of new branches, rendering this universal arborescence of complexity 'alive'.

The work starts with mathematical analysis of the simplest physical systems introducing the new paradigm and representing the lowest levels of the universal complexity; then the obtained description is extended to the arbitrary complex behaviour, and finally some applications to systems from higher levels of complexity are outlined, including the highest levels of human activity directly involving the forthcoming global transition of the Revolution of Complexity.

Whereas further development of various particular directions of the science of complexity is certainly implied by the necessarily brief outline presented in this work, we emphasize the irreducible conceptual basis of the Universal Science of Complexity, proposed in the form of the dynamic redundance paradigm together with its main consequences, and their fundamental uniqueness.

That is the Way from the End to the Beginning.

# Universal Science of Complexity: Main Results

This work introduces and develops a new, universal method of the conscious comprehension of reality providing the objective, completely causal and exhaustive description of the world in general and any its particular phenomenon (Kirilyuk 1997). The method is based on the naturally appearing concept of the *universal dynamic complexity* which is applied to the rigorous analysis of dynamical systems of various origins and levels. In this way one obtains the basic outline of the *universal science of complexity*, after a number of recent unsuccessful attempts to create it (Horgan 1995). This nontrivial result is possible only due to the new interpretation of the main dynamic equations that does not use any intuitive assumptions, or artificial additions to the existing formalism, but naturally *extends* it to its full meaning, largely reduced in the canonical interpretation by application of various versions of 'perturbation theory'.

Namely, it is shown that if one avoids, in a universally applicable fashion, the usual simplifications of perturbation theory, then practically *any* dynamic equation describing a nontrivial interaction process with more than one effective dimension demonstrates *fundamental dynamical splitting* into many equally possible and qualitatively similar, but quantitatively different, 'integrable' (analytically tractable) versions, each of them corresponding to a 'possibility', called *realisation*, for the real process in question. Only one of those possibilities/realisations can actually appear at any particular moment, since each of them is generally equivalent to the ordinary 'complete' solution, and therefore the realisations are definitely *incompatible* among them forming a (dynamically) *redundant* set. This leads to a quite new picture of reality where it is obtained as a hierarchy of *possible* states (realisations) permanently *replacing* one another at various scales and in different regimes, which gives the observed variety of always intrinsically unstable, *probabilistic* forms and types of behaviour. Probabilistic realisation 'switching' gives rise also to the property of *time* characterising the rate of their change (see also below).

This picture reveals the fundamental origin, and meaning, of *randomness* in the world emerging as a purely dynamic, deterministically based phenomenon of redundant multivaluedness, which provides the causal dynamic foundation for the concept of *probability*, and a practical method of theoretical calculation of probabilities, in principle, in arbitrary case. Despite any external illusions of stability and order, the world is definitely represented now as a process of permanent spontaneous (i. e. essentially *unpredictable* and *irreversible*) change, both in every its part and as a whole; it is only (chaotic) *change* that can be *permanent*, and only randomness that can be ordered. In fact, the extended notion of *dynamical chaos*, applicable to any kind of system and expressing the complexity of its dynamics, is just equivalent to that *partially ordered randomness* of purely dynamic origin.

These fundamental conclusions are rigorously deduced from the first-principle, causal, and technically simple analysis of dynamic equations of a very general form, and also demonstrated for the most important particular cases, such as the Schroedinger equation in quantum mechanics (Kirilyuk 1995a,b, 1996). The resulting 'method of effective dynamical functions' (Kirilyuk 1992) is a generalisation of the elementary 'method of substitution of variables' in the form of the well-known 'optical potential method' (e. g. Dederichs (1972)), but avoiding the usual perturbative reduction of the latter which just 'kills' all the complexity-multivaluedness. In the full, complex-dynamical version of the method the key property of dynamic redundance appears as a result of the *essentially* nonlinear, multivalued dependence of the effective interaction potential on the eigenvalues to be found which corresponds, physically, to universally defined self-sustained feedback 'loops' of the unreduced interaction processes, naturally expressing the fact that 'everything depends on everything' and leading to the global instability with respect to redundant realisation formation (see also below). This essential dependence of the potential on the quantities to be found is absent in the ordinary truncated, perturbative formulations, where the unknown eigenvalues occur only at one, canonical position, which leads to a single-valued (effectively one-dimensional) solution. The crucial qualitative advance of our approach with respect to the ordinary perturbational reduction is attained due to a particular combination of the *exact* expression of those *essential* dependences responsible for dynamical splitting into multiple realisations with an unavoidable approximate estimate of qualitatively less important relations (they also play an irreducible role for further development of the secondary, fine structure of complexity in the form of its fractal structure, see below).

We call this new concept *dynamic redundance paradigm* (or fundamental multivaluedness of dynamical functions (FMDF), or fundamental dynamic uncertainty). It is equivalent to the concept of *universal dynamic complexity*, since the latter is naturally determined by a strictly increasing function of the number of realisations, equal to zero in the exceptional case when a system has only one realisation, which means also that the system is regular. It is only that, actually pathological, kind of system and behaviour which is exclusively described by the canonical, or *linear* (mechanistic), or *unitary* science, dismissing all the realisations of a system but one by the explicit or implicit reduction within a version of perturbation theory (irrespective of any superficially defined algebraic or geometric 'nonlinearity' and empirically fixed randomness).

In particular, practically all the previous attempts to introduce (dynamic) randomness and complexity have been performed within the single-valued paradigm of the canonical science, and therefore are inevitably reduced to their artificial insertion à la *deus ex machina*, e. g. in the form of basically unknown and indefinite 'influences' of either external 'environment', or internal 'deeper levels' of reality formally described as ambiguous 'nonlinearity', 'decoherence', or artificial 'coarse-graining' (see also below). One should clearly distinguish that kind of 'scientifically looking', often pompously advertised and technically 'sophisticated', but logically elementary, trickery and imitation around 'science of complexity' from its causally irreducible, logically consistent and completely transparent introduction which is absolutely unavoidable for such a qualitatively new level of comprehension of reality, as it is demonstrated once more by the dynamic multivaluedness paradigm.

The new paradigm universally resolves also the problem of *(non)integrability*, since it becomes clear that every regular, effectively one-dimensional case with one, and only one realisation corresponds to an integrable problem, while the multiplicity of realisations for a generic problem defines and explains its 'nonintegrability': one cannot obtain one solution for a problem definitely having many of them. In return, now *every* problem can be *actually integrated*, i. e. *completely solved*, by simply applying the universal nonperturbational method of the unreduced science of complexity that uses the 'effective dynamical functions' and represents the extended version of the ordinary 'method of substitution' (Kirilyuk 1992, 1995a,b, 1996). The 'price' to pay for this privilege to solve any problem is the fundamental dynamic multivaluedness of the obtained solution that contains much of internal intricacy (described below) and conceptual novelty.

In other words, the *complete (general) solution* for *any* problem can indeed be obtained, but only as a multitude of elementary solution-realisations, each of them being roughly equivalent to the ordinary, linear-science, effectively one-dimensional 'complete' solution. This explains the *generic* failure (divergence) of perturbational approaches: they try as if to make their expansions around the omnipresent singularity of 'branch point', since the dynamic multivaluedness corresponds to 'branching at every point'. In particular, the permanent global, chaotic jumps of a system between the quasi-regular realisations cannot be described by perturbation theory in principle (they *seem* to be *infinitely* sharp and rapid, within the linear science analysis), whereas they are explicitly derived, as a major, basically continuous (causal) dynamical process, within the general version of our formalism applied to the well-known equations (see below).

For the same fundamental reasons, all the conventional concepts of the linear science around dynamically complex behaviour, based invariably on perturbative approaches, are profoundly deficient, both conceptually and formally. Thus, the popular reduction of dynamical chaos to the 'exponentially divergent' trajectories or other 'states', as well as similar theories based on a particular *mathematical* dependence as a definite *signature* of a *qualitatively* new reality, turn out to be basically wrong, since the dynamics of any complex system is certainly *not* a uniform (unitary, or single-valued) evolution, whatever is its formal law, but rather a 'random walk' of Brownian type, with the chaotic jumps between the incompatible state-realisations driven by the intrinsic *global instability of redundance* (induced eventually by any participating interaction, if it is not pathologically simple). It is not difficult to show that the states of a system and their chaotic sequences (generalised trajectories) in reality diverge, in average, according to a power-law dependence related to the first term in the exponential-dependence expansion in a series which is actually fictitious, however, since it can never be 'summed up' in the real system evolution because of the causally *random* jumps between the generally *incoherent* realisations occurring just when the argument is around unity, the critical value for the series convergence. Correspondingly, the canonical formalism of 'Lyapunov exponents', based on the incorrect extension of the perturbative, *locally* valid linearisation around *one* particular point to the *global* system behaviour (see e. g. Lichtenberg and Lieberman (1983), Schuster (1984), Zaslavsky (1985), Zaslavsky *et al.* (1991), Ott (1993), Chirikov (1995a,b)) gives essentially wrong, qualitatively misleading results, even though it can sometimes roughly *simulate* the global instability. We show that the same profound non-uniformity of (complex) dynamics of practically any real system actually invalidates other canonical, 'well-established' exponential dependencies of the single-valued, linear science, such as those in the unitary 'evolution operator' and Feynman 'path integral'. Their respective complex-dynamical extensions contain probabilistic sequences of basically linear (eventually, power-law) dependencies involving the generalised Lagrangian.

In general, the single-valued, unitary paradigm of the canonical, essentially linear science is nothing but a one-dimensional *projection* of the unreduced multivalued reality and correspondingly includes all possible *simulations of complexity* depending on the projection view and taking the form of external and always somewhere incomplete, mutually separated *signatures, 'fingerprints', or shadows of complexity* defined only *formally* and ambiguously as 'nonlinearity', 'chaoticity', 'self-organisation', 'adaptability', 'fractality', '(self-organised) criticality', etc. (see e. g. Prigogine and Stengers (1984), Babloyantz (1986), Nicolis and Prigogine (1989), Prigogine (1995), Haken (1988), (1996), Haken and Mikhailov (1993), Gaponov-Grekhov and Rabinovich (1990), Peintgen, Juergens, and Saupe (1992), Allen and Phang (1993), Coveney and Highfield (1995), Bak (1996)), which determines the basic origin of the evident failure of this canonical, linear approach to reproduce the *natural unity* of being within a single concept of dynamic complexity (Horgan 1995, Bricmont 1995). Moreover, even each particular type of the mechanistic simulation of complexity is evidently inconsistent: the true randomness cannot really be reduced to an 'involved regularity' or a special mathematical function ('exact solution'), intrinsic 'noncomputability' cannot be simulated by a regular computation (cf. Penrose (1994)), and *self*-organisation should involve an irreducible autonomous 'emergence' of a structure, rather than its implicit insertion 'by hands'.

Contrary to this basic limitation of the linear-science projection of reality, the universal science of complexity, based on the dynamic redundance paradigm, correctly reproduces within a *single* approach *all* the existing cases of complex behaviour with its full diversity and inherent *constructive* contradictions (dualities), actually forming the irreducible *basis* of dynamic complexity as such. The externally different, and even opposed, types of behaviour, reduced mainly to various combinations of the entangled order (regularity) and structured disorder (randomness), are explicitly shown to be profoundly connected by the unique dynamic complexity that simply manifests its different aspects in each particular case, depending on the respective, partially unpredictable and dynamically determined, 'proportions' of regularity and randomness. Thus, a characteristic regime of 'dynamical chaos' (in the narrow sense) is obtained when the system parameters are such that realisations are noticeably different and chaotic transitions between them are relatively frequent, whereas in the opposite case one obtains the regime of (generalised) 'self-organised criticality', or 'structure formation', when the realisations are grouped into closely spaced ensembles of similar members, and chaotic jumps between the groups are relatively rare. Therefore any 'self-organised', more stable structure or dynamical regime is still impossible without the internal true chaoticity (randomness), while chaos always involves some regular structures (corresponding to individual realisations and a nonuniform distribution of their probabilities). In particular, it becomes clear that various, now intensively discussed procedures of 'control of chaos' are reduced to the generalised self-organised criticality and therefore will practically always contain some remnants of randomness (dynamic unpredictability). The reverse is also true: any 'self-organised' behaviour can be considered as an intrinsic control-of-chaos regime, where the 'control' is 'performed' by the system itself. The extended 'self-organised' type of behaviour can also be considered as a generalised 'bound motion', eventually giving rise to all localised, 'distinct' type of objects and dynamic regimes, while the highly irregular, 'uniform' chaos is the universal prototype of all the 'free motion', 'distributed' and widely 'extended' states, irrespective of explicit appearance of this internal chaotic structure of the bound and free motion states.

Since fundamental dynamical splitting of reality, described as its redundance and giving rise to complex behaviour, naturally forms hierarchically breeding levels, the dynamic complexity of the world follows the same tendency and 'spontaneously' *emerges* as a universal hierarchy of qualitatively different observable objects and types of behaviour. Such *hierarchical structure of complexity* manifests itself at all scales, giving rise to the complex-dynamical extension of the notion of *fractal*, acquiring now basically *dynamic* and causally *probabilistic* character that largely exceeds the simplified mechanistic 'self-similarity' (or 'scale invariance') of the canonical notion (Mandelbrot 1982, Feder 1988, Peintgen, Juergens, and Saupe 1992, Nakayama, Yakubo, and Orbach 1994) always demonstrating the same purely formal origin that avoids any *irreducible* relation to reality and its *direct, adequate* mathematical representation. The dynamical fractal of the universal science of complexity is directly obtained as the general *solution* of a main dynamic equation describing the problem in question. It appears as 'non-stop' dynamical splitting of a problem governed always by the *same* FMDF mechanism and *explicitly* realising the intrinsic *unification*, physical and conceptual, of the *causally probabilistic*, and therefore *self-developing* ('living'), 'time-generating' world dynamics. The intricacy of the resulting *fundamental dynamical fractal* (of a problem or even the whole world) demonstrates the fully developed consequences of the *universal* 'eventual' integrability of *every* problem attained within the dynamic redundance paradigm. The obtained problem solution, as well as reality it describes, has the 'infinitely' *entangled* structure, which corresponds to *nonseparability* (physical and mathematical) of a problem. This means, in particular, that the causal nonseparability-entanglement of the dynamical fractal always unpredictably changes (develops itself) in its details. The fundamental dynamical fractal thus obtained is the *unified* extension of various simulations of fractal branching in the linear-science 'complexology', such as 'bifurcations' ('catastrophes'), 'strange attractors' (including the latest variations around 'bubbling attractors', 'riddled' or 'intermingled basins of attraction', etc.), 'Cantor-like' sets, 'fractal spectra', 'intermittency', various types of 'mixing', 'entanglement', and 'percolation', together with any other possible inhomogeneities: there is no structure in the existing reality other than the self-developing universal dynamic fractality. Therefore the fundamental dynamic fractal of the world, as well as its mathematical presentation within the dynamic redundance paradigm, naturally *produces* not only 'fractal-looking', entangled 'networks' of smaller elements, but also quite 'distinct', quasi-regular, 'separate' objects like an egg or a smooth stone.

From the other hand, the 'distributed' regimes of complex dynamics referred to as 'turbulence' and 'self-organised criticality' and poorly understood in the linear science also obtain now a well-specified interpretation as 'essentially multilevel' regimes of complex behaviour in which there are many closely spaced levels of complexity connected by a dense fractal network of 'sublevels', so that the transitions between realisations from many hierarchically organised levels contribute to the observed patterns and cannot be easily separated.

The manifestations of the universal dynamic complexity are studied in more detail for a number of particular and general cases. The lower levels of complexity correspond to the *elementary fields and particles* conventionally (and inconsistently) described within such linear-science domains as quantum mechanics, field theory, particle physics, (quantum) theory of gravitation, and cosmology. Their considerably extended versions are naturally unified in the *single* theory of the *quantum field mechanics* corresponding to specification of the universal hierarchy of complexity for several its lowest levels.

At the lowest observable level of elementary particles, we obtain the essentially complete realisation of the unreduced version of the *causal wave mechanics of Louis de Broglie* (see e. g. de Broglie (1956), (1964), (1971), (1976), Andrade e Silva (1973), Lochak (1973)) known as the *double solution* and *hidden thermodynamics of the isolated particle* and extended now by the natural inclusion of the dynamical chaos concept. This *double solution with chaos* presents an elementary fermion, e. g. electron, as a permanent *process* of *quantum beat* consisting of unceasing series of temporally periodic cycles of spatially chaotic *reduction* (a self-amplified, effectively nonlinear *squeeze* to a small volume) and the reverse *extension* of *electromagnetic* (proto)field, induced *exclusively* by its *homogeneous* attractive *coupling* to a *gravitational* 'background' medium (or 'protofield'). Such specific behaviour of permanent chaotic quantum beat is rigorously obtained from the equation of a very general form, describing this fundamental electro-gravitational coupling, only due to the extended interpretation of the fundamental dynamic multivaluedness, whereas the ordinary, single-valued analysis of the linear science would predict in this case a single, more or less homogeneous, 'fall' of the electromagnetic field onto the gravitational background. The frequency, $\nu$, of the quantum beat is very high (of the order of $10^{20}$ Hz for the electron) and characterises the basic velocity of the 'time flow' giving the elementary time unit of the world, $\Delta t = 1/\nu$; it is determined eventually by the magnitude of the driving electro-gravitational coupling and can be expressed through the mass, $m$, of the elementary field-particle, in accord with a basic relation proposed by de Broglie ($h\nu = mc^2$, where $h$ stands for Planck's constant, $c$ is the velocity of light) and expressing the fundamental connection between dynamic complexity, mass, energy and time in our approach (see also below).

The obtained complex-dynamical solution of the quantum field mechanics reproduces all the previously known and expected basic properties of de Broglie's double solution and the corresponding *causal and complete version of quantum mechanics*, thus directly and *causally* explaining the observed peculiar properties of the *physically real* fundamental entities (fields). In particular, it completely unveils the origin of the most 'mysterious' properties of 'wave-particle duality' (de Broglian 'material wave-particle') and 'quantum indeterminacy' (de Broglian 'hidden thermodynamics' of a particle) by showing that they are the inevitable, standard manifestations of the dynamically complex behaviour of interacting fields. The unreduced process of attractive interaction between the primal electromagnetic and gravitational media (protofields) leads, through the universal and physically transparent feedback mechanism, to the global dynamic instability with respect to auto-squeeze of the extended electromagnetic field to a small volume (which is the 'objective, spontaneous reduction/collapse' of the elementary field). This transient, very short-living squeezed state is the 'particle', or 'corpuscular' stage, and aspect, of the fermionic field and 'wave-particle duality'. Due to existence of other similar, i. e. dynamically redundant, squeezed states at different possible centres of reduction and one extended state of the field, this virtual 'particle' is unstable and quickly *transformed* first to the extended state, which realises the 'wave' aspect of duality, and then again to the squeezed, corpuscular state at *another* centre of reduction, 'chosen' by the system necessarily *at random* (probabilistically), which demonstrates the dynamic origin of 'quantum indeterminacy' (it is thus a manifestation of the *same* universal dynamic uncertainty of any complex behaviour). Therefore there is no any antagonistic rupture between the field (wave), and the particle (corpuscle), since any elementary object is present *alternatively* and successively in *both* these qualities, and the corresponding *real* states, during the phases of extension and reduction respectively. In fact, these dualistically opposed states need each other in order to maintain the very existence of a system with interaction, which provides the decisive clarification and extended interpretation of the famous idea of 'complementarity' of Niels Bohr who seemed to have intuitively 'guessed' this intrinsic property of the hidden complex dynamics of the micro-objects, but remaining always honest in his relations with the truth was obliged to recognise the impossibility of explicit discovery of its causal origin within the existing unitary paradigm (contrary to certain modern imitations of creation, see also below).

The *dynamic*, self-sustained *discreteness* of the quantum-beat process can be related to fundamental *globality* (wholeness) of (any) complex dynamics (see below) and is at the origin of that special, properly *'quantum'* behaviour of micro-objects which is *simulated* by the formally postulated 'quantization rules' in the conventional approach; in the quantum field mechanics we causally *obtain* the real, *physical quantization process* taking the form of the quantum beat dynamics. Correspondingly, Planck's constant $h$ equals to the mechanical action integral taken over one period of quantum beat (which is fundamentally different, however, from any linear, or even anharmonic, 'oscillation'); its value is determined eventually by the parameters of the driving electro-gravitational interaction, while the mechanical action is a universal measure of dynamic complexity (see also below).

The mathematical 'wavefunction' of the ordinary quantum mechanics acquires now a fully realistic interpretation in which it is represented by the sequence of the extended, intrinsically *chaotic* and *transient* states of a *physically real* (basically electromagnetic) field between the causal reductions (due to its coupling to gravity), possessing dynamically maintained temporal and *partial* spatial coherence. The latter provides also the realistic, causally specified version of the observed 'de Broglie matter wave': it is a superposition of many spatially coherent structures in the extended state of the globally moving fermionic field resembling a standing wave, while the underlying process of quantum beat includes also a specified, non-zero proportion of cycles of reduction-extension with spatially irregular field structure in the extended phase and distribution of the reduction centres (this 'proportion of irregularity' attains hundred per cent for the elementary fermion at rest and tends to zero when the particle velocity tends to that of light, see also below). The 'phase accord theorem', used by de Broglie (1924) to deduce his famous expression for the wavelength of a moving particle, also acquires its causal completion: the 'internal clock' of a particle 'transported' by the 'associated' wave remains always in phase with that wave simply because both motions are in reality different aspects of one and the same quantum beat process in which the particle-clock (the squeezed corpuscular state) is permanently transformed into the 'transporting' wave and back, with a probability distribution of the directions of chaotic 'jumps' of the particle between the reduction centers that corresponds to *average* displacement of the particle with the given velocity.

Note that the proposed extension of the unreduced version of the double solution, as well as the original version itself, should be clearly distinguished from the considerably simplified, actually mechanistic version of 'Bohmian mechanics' or 'quantum theory of motion' (see e. g. Holland (1995), Berndl *et al.* (1995), Goldstein (1998)). The latter was originally proposed by de Broglie himself under the name of 'pilot-wave interpretation' as an explicitly simplified, practically oriented mathematical scheme of the detailed realistic picture which escaped the rigorous, noncontradictory description. Whereas de Broglie later concentrated his efforts on the search for such consistent realistic description, the truncated pilot-wave version was 'rediscovered' 25 years later (in 1952) by David Bohm and is especially actively promoted by his followers in the last period as a 'realistic' and fully consistent 'interpretation' of quantum mechanics, with vanishing reference to the unreduced original version of the realistic approach. However, it is easy to see that purely 'interpretational' approach of Bohmian mechanics, actually reduced to a mathematical reformulation of the Schroedinger equation without any qualitative novelty (like the dynamic redundance or double solution), reproduces all the main difficulties and basic incompleteness of the standard interpretation, now in their 'reformulated', and characteristically less transparent, version (cf. d'Espagnat (1994), (1998)). Thus the 'classical particle' of an unspecified physical origin and structure is introduced simply by a formal postulate and the fundamental source of randomness remains ambiguous, apart from many other particular difficulties. The whole approach suffers from the mathematically postulated character of the essential concepts and entities deprived from the directly specified physical contents, quite similar to any other mechanistic 'interpretation of quantum mechanics' leaving it as mysterious as it is explicitly and consistently acknowledged in the most transparent standard (Copenhagen) interpretation.

The realistic approach of the quantum field mechanics provides complex-dynamical, physically and mathematically complete extensions of all the known basic entities and their properties, and among them the causal, universal interpretation of *mass-energy* as the temporal rate (intensity) of the spatially chaotic quantum beat of the coupled electromagnetic and gravitational protofields which possesses all the necessary properties of *inertial and gravitational mass* being thus *equivalent* among them par excellence, up to a subjective choice of the participating coefficients (see below). It is shown that the mass-energy so defined is a measure of dynamic complexity of the elementary field-particle, and therefore the property of *inertia* and conservation of mass-energy are but particular manifestations of the single fundamental principle of nature, the complexity conservation law (see also below). Physically, any massive object 'resists' to an attempt to change its motion simply because it is already in a state of irreducible internal 'thermal' motion (chaotic quantum beat), which is similar to resistance to compression of a gas of chaotically moving molecules.

Not only the quantum field mechanics naturally unifies inertial and gravitational manifestations of mass, but it realises the full intrinsic unification of quantum mechanics, 'relativity', gravity, 'field theory', and 'particle physics' considerably extending and causally demystifying each of them.

Thus, the *universal gravitation* is seen as a result of the same process of quantum beat, since it is nothing else than dynamic attraction of any elementary particle, and therefore any macroscopic body, to the *unique* continuous gravitational background (protofield) that has its own 'elastic' reaction to the reduction events for each particle-process inevitably felt by other 'subjects' of reduction (which explains the above causal 'equivalence' between the gravitational and inertial aspects of mass). It is very important here that the *whole* universe is produced by coupling of the *same* two protofields, electromagnetic and gravitational media, which provides the necessary *physical* basis for unity of the world that can be described now as a kind of (three-dimensional) 'sandwich' made from the two protofields with a help of coupling interaction(s) among them, while the FMDF mechanism assures the sufficient diversity of the dynamical structure and behaviour naturally *emerging* from this simple starting configuration. The intrinsically *universal* gravitation thus obtained is not only 'unified' with quantum mechanics, but *inseparably* related to it by their common origin: *any* gravitational attraction is essentially a discretely structured, non-stationary, *quantum process* (though normally with a quasi-classically large number of quantized pulling events), since it is driven by the quantum-beat pulsation, just determining the specific, dynamically discrete, or 'quantum', behaviour of an a priori uniform and stationary system of interacting protofields. The intrinsically discrete, quantum character of gravitation will explicitly manifest itself only at the (modified) Planckian-unit scale (see below). This interpretation of (quantum) gravitation and other fundamental properties of the world (see also below) gives rise to the unique self-consistent resolution of the old 'action-at-a-distance' mystery by revealing a causally continuous sequence of really occurring quasi-local events behind any apparent 'nonlocality' of interactions, quantum effects, etc.

Although in the macroscopic, classical limit this intrinsically quantum gravity of the science of complexity produces in many cases the same measurable results as those predicted by the Einsteinian 'general relativity', their interpretation and a number of important particular applications follow a dramatic extension towards causal consistency. Gravitation and its influence are not characterised any more by an ambiguous, purely abstract 'deformation of (mathematical, four-dimensional) space and time' - which is evidently not more realistic than the canonical interpretation of the wave-particle duality, so definitely refuted by Einstein himself - but by the *local* 'density', or 'tension', of the dynamically varying, physically realistic gravitational medium (always in connection to the coupled and directly perceived electromagnetic medium). The inhomogeneous macroscopic distribution of this tension, alias 'gravitational potential', can indeed be presented as *local* deformation of the elastic 'gravitational medium' around massive bodies, and the *local* dynamics of this system can indeed be described as the uniform motion in the curvilinear system of coordinates coinciding with the 'lines of equal tension' (geodesics), but this is *only one* of the many possible ways (coordinate systems) of the *formal* description that *cannot* be identified with the unique objective reality and can neither be applied, even formally, to the *global* (averaged) dynamics of the whole universe. The objectively unique, and in this sense *absolute*, reality does exist and corresponds, for a 'normal', self-sufficient universe, to the *globally flat* space which can be effectively 'deformed' (in the above realistic sense of the nonuniform gravitational medium tension) only *locally*, around higher concentrations of the dynamically appearing fermionic reduction events. As to the particularly ambiguous concept of the 'curved time' of the standard relativity mysteriously 'mixed' with 'curved space' (whereas one fails to explain even what the *real, irreversible* time and space *is*, apart from a mathematical 'coordinate'), it actually tries to simulate the complex dynamics of the elementary events (quantum-beat reductions) whose rate of appearance, determining the causal time flow (see also below), is realistically and transparently related to the local gravitational tension (potential). It is evident that the complex-dynamical, causally substantiated picture of the globally flat, but internally structured, probabilistically developing ('living') universe changes dramatically the conventional cosmological applications of the standard general relativity around simplistic single-valued 'solutions' for the whole universe à la Friedmann (already because they are based on the incorrect mechanistic extension of the *locally* possible geometric interpretation of gravity to the *global* scale), including the 'open' or 'closed' universe alternative, the idea of the 'explosive' expansion of the universe from a point-like initial state (Big Bang, inflation), the problems of 'quantum cosmology', etc. The revised, complex-dynamical cosmology naturally emerges in its (causally) *quantum* version which provides, in particular, a consistent solution to the well-known 'problem of the wavefunction of the universe' based on hierarchical *local* splitting into dynamically redundant, *incompatible* (non-superposable), probabilistically *changing* (and thus *time-generating*) structures/events. It is clear also that all the existing simulations of the causal quantum dynamics of gravity within the same purely abstract approach of the mechanistic science, reducing it to various formal constructions basically detached from reality, such as 'special', e. g. 'noncommutative', geometries or topologies, are condemned to failure by their zero-complexity, irreducibly 'non-dynamical' origin.

The same unified picture of the quantum field mechanics provides the causal, physically transparent notions of *space and time* crucially extending the reduced, formal notions of the canonical science including those from 'special and general relativity' and the 'new physics' in general. In the science of complexity, *time* characterises the intensity (rate, or frequency) of spontaneous and probabilistic realisation emergence (constituting the generalised *event*) and is intrinsically *irreversible* (which reflects its profound relation to the dynamical randomness), while *space* defines the global (structural) quality of the emerging realisation perceived largely as its 'dimension' (size) and is naturally *discrete*. Space has a well-defined *material*, tangible basis, it is not only the structure of matter, it is matter itself, since matter appears only as structured entities (realisations), in full agreement with the famous Cartesian concept of *étendue* (*res extensa*). Time is not a material entity, it is rather a *sign* that matter-space exists, and especially that its new (spatial) form-realisations *emerge* as a sequence of incompatible and therefore partially *unpredictable* events; time emphasises the irreducibly changeable, unstable (multivalued) and probabilistic character of being revealed by the science of complexity (we can say also that different *moments* of our *causally emerging* time of the *single* real world replace different *versions* of the *whole* world ambiguously *coexisting* in an indefinite 'set of worlds' introduced *axiomatically* in the well-known 'many-worlds interpretations'). Therefore time cannot be considered as a *real* 'degree of freedom', or 'dimension', of a world completing its 'spatial dimensions', as it is postulated in the canonical 'relativistic' paradigm of the linear science referring to a purely formal, and actually reduced, symmetry (in fact, we show that the 'principle of relativity' and the related formal symmetries of the canonical science simulate the same universal symmetry of complexity, expressed as its conservation, see also below). Space and time can be mutually 'entangled' only in the indirect sense of dualistic 'dispersion relation' among them expressing the elementary dynamics of a level of complexity. Space means 'what' (the existing object is) and corresponds to a noun, time shows 'how' (it exists) and corresponds to a verb, and any formal, mechanistic 'equivalence', or 'symmetry' between these two basic aspects of complexity can only be extremely superficial (they are rather *dualistically* opposed, or complementary, to each other providing thus another manifestation of the extended principle of complementarity). This causal definition of space and time and the fundamental difference among them are also reflected in the direct relation between space and the emerging dynamic complexity-entropy, from one hand, and of time with the disappearing dynamic complexity-information, from another hand (see below for definitions of the generalised entropy and information).

Hierarchical structure of complexity determines the hierarchical structure of space and time represented by a fractal network of dynamically related, 'breeding' levels. In particular, any dynamically complex behaviour (resulting actually from any nontrivial interaction) involves permanent realisation change, and therefore cannot provide a truly static and regular, time-independent regime, which naturally gives rise to the 'time flow' of the corresponding level. Most important is the lowest level of the hierarchy formed by the quantum-beat dynamics; the latter causally 'produces' those 'embedding' space and time of our world which usually appear to be perfectly uniform at the higher, 'classical' levels of complex behaviour, but are irreducibly and 'spontaneously' quantized at their basis, as should always be the case for *any* real entity described by the universal science of complexity. In particular, the 'tissue' of the physically real space in which we live and which is indispensable for *any* existence is permanently 'woven' from the electromagnetic protofield over the hidden gravitational matrix-protofield by the reduction-extension events of the fundamental quantum-beat process, thus playing the role of a complex-dynamical 'loom' (its peculiarity is that the produced fundamental 'fabric' of space immediately returns to the initial 'unwoven' state, but then is again quickly 'rewoven' by the 'loom', in a dynamically continuous and spatially chaotic regime). Note in this respect that the starting equations of the quantum field mechanics do not contain any *a priory* space and time variables, nor any discreteness or irreversibility, they are explicitly *obtained*, within a rigorous mathematical description, as manifestations of the complex behaviour emerging due to the dynamic redundancy. Another fundamental property of the world which is rigorously obtained in the same causal fashion is the number of its spatial 'dimensions' (degrees of freedom). Indeed, according to the basic complexity conservation law, the number of the degrees of freedom remains unchanged in the closed system with interaction, but they normally change their 'quality' because of the complex-dynamical entanglement (see also below). Therefore the two protofields and their interaction (coupling) can give three and only three spatial degrees of freedom and one time variable characterising the three-dimensional event appearance.

The same description reveals also the extended, causal origin of the effects of *special relativity* of thus emerging time and space which are interpreted as 'transport effects' of the quantum-beat dynamics, accompanying the global transport of probability of reduction, and thus of the appearance of the 'corpuscular' state, for the moving field-particle. The *state of motion* itself can be rigorously defined now as a state with complexity-energy exceeding the well-defined minimum value attained for the case of completely uniform distribution of reduction probability which corresponds to the *state of rest* of the field-particle (the state of rest at a higher level of complexity is also defined as the one with locally minimal dynamic complexity, and respectively for the state of motion). The famous 'time retardation' relation of the canonical relativity is correctly reproduced and completely demystified now, together with the notion of time itself: when one is 'at the moving edge' of a currently emerging structure (e. g. that of a moving field-particle) time formally stops in that 'reference frame' (and at that level of complexity) simply because the elementary increment of *local* time totally *results* from the *relative* motion of the edge. This causal 'special relativity' is thus intrinsically unified with the causal quantum dynamics, as opposed to their mechanistic superposition, in the form of the respective 'mysterious' postulates, within the canonical scheme of justification of the Dirac equation. The effects of *general relativity* of time and space are provided with the same physical transparency (see above) and naturally (dynamically) *unified* now with those from the (extended) special relativity and quantum dynamics, contrary to the canonical versions of special and general relativity whose basic separation is hidden in the axiomatically imposed, abstract 'principles' of relativity, etc. Actually 'relative' is always the level of dynamic complexity at which the 'observer' (represented by another complex system) is placed, and the level of complexity is characterised by the temporal or spatial rate of event (realisation) emergence which is always registered *with respect to* similar rate from another complex-dynamical process (attributed to the observer). This description and interpretation of relativity are universally applicable at any level of complexity, providing, for example, an objective explanation for relative, or 'subjective', time flow even at the level of human 'psychological' time perception (the subjectively perceived 'time flow' effectively 'slows down' when there are many big, intrinsically 'striking' events, etc.).

Many other basic properties of elementary particles and the corresponding relations are provided with transparent interpretations within the quantum field mechanics that forms therefore the basis for the new, causal description of microworld which should totally replace the fundamentally reduced, and therefore unrealistic and separated, versions of the linear science.

In particular, the property of spin escaping any realistic interpretation within the linear science, naturally emerges now as a highly nonlinear vortex motion of the elementary field in the stage of self-sustained reduction driven, in principle, by the same shear instability as the motion of water flowing from a basin through a small hole in its bottom; the key novelty of the quantum field mechanics is that it provides the purely dynamic origin of permanent 'spontaneous' emergence of those 'holes' in the formally uniform system of interacting protofields. The same vorticity, though rather out of the maximum-squeeze stage of quantum beat, accounts for the causal origin of the magnetic field (while the 'oscillatory' component of the quantum-beat process gives rise to the electric component of the electromagnetic field). We arrive thus to another manifestation of the intrinsic (dynamical) unification of the quantum field mechanics which includes this time spin, magnetic and electrical fields, and gravitation; we can say that our causally interpreted spin-vorticity (as well as the related magnetic and electric fields) is a 'gravitational effect', in the sense that it is driven by the fundamental coupling to the gravitational 'protofield'.

Photons correspond to propagating 'quasi-free' perturbations of the electromagnetic protofield emitted by a quantum-beat process (and further re-absorbed by another particle-process), and their discreteness is totally due to the quantized character of the quantum beat. The photon does not possess its own mass because it does not possess any internal complex dynamics; it is indeed rather close to the canonical 'linear wave', even though in real interaction processes the photons often 'inherit' the features of complexity from the related dynamics of the emitting and absorbing massive field-particles (further refinement of this picture involves a possibility of spatially *regular* quantum beat in the isolated photon dynamics, which could help to complete the description without changing the essential points). The quantum-beat process itself can be considered as unceasing cycles of emission and absorption of many transient, or 'virtual', photons, whereas the interaction between two particles is indeed represented by the exchange of transient (but real!) photons, the process which can now be presented in its physically transparent, causal version, contrary to the purely mathematical, and irreducibly perturbative, simulation of the canonical 'quantum field theory'.

Correspondingly, the property of electric charge also gains its causally complete interpretation as another expression of the dynamic complexity of the quantum-beat process. This explains the discreteness of the elementary charge, $e$, and proportionality of its square to Planck's constant $h$ also expressing dynamic complexity per cycle of quantum beat, as mentioned above (we refer here to the well-known relation $e^2 = \alpha \hbar c$, where $\alpha = 1/137$ is the 'fine structure constant' and $\hbar = h/2\pi$). The fact that there are two and only two 'opposite in sign' types of electric charge with the same absolute elementary value is due to the 'two-phase' structure of the elementary cycle of the fermionic quantum beat, which leads to subdivision of all the observable quantum-beat processes (corresponding to individual massive elementary fermions) into two classes with the opposite temporal phases of pulsation, while in each class the quantum-beat oscillations are in phase (generally similar to the behaviour of mechanical systems producing 'oscillons', see Umbanhowar, Melo, and Swinney (1996)). Therefore the elementary particle charge is a somewhat more direct expression of the observed quantum-beat complexity than the rest mass or Planck's constant which can be related also to the hidden parts of the quantum-beat dynamics. In any case, conservation of charge emerges as another manifestation of the universal complexity conservation law.

The above picture shows also that there is the profound physical coherence in the universe dynamics: the most fundamental level of the world dynamics is characterised by the exact temporal coherence of all the existing quantum-beat oscillations, up to the half-period shift between the particles with unlike charges (otherwise we would have more than one absolute value of the measurable elementary charge); the universe has therefore its unique most precise and physically real, distributed 'rhythm' perceived 'simultaneously' in every its location (by contrast to this temporal coherence, there is no general spatial coherence between the quantum-beat oscillations). This *absolute* aspect of time is a manifestation of the same unity in the world construction that explains the universal character of gravitation and definitely rejects the exaggerated, superficial 'relativism' of the unitary, simulative 'new physics' (cf. Davies (1989)). Another manifestation of this unity directly involved with the above causal interpretation of the electromagnetic interaction and the property of (quantized) electric charge is the naturally attained unification of electromagnetic and gravitational phenomena, so intensively and unsuccessfully sought for by the canonical, unitary science (in addition, both types of phenomena are obtained in their most complete, quantum interpretation). We clearly see now that the observed electromagnetic and gravitational interactions are different but related aspects of the same quantum-beat process physically 'transmitted' through the two coupled sides of the 'world sandwich', the electromagnetic and gravitational proto-media respectively. This intrinsic connection between electromagnetism and gravitation is expressed, in particular, by the (modified) Planckian units (see below).

The total dynamic construction of the real, many-particle world is a natural extension of the quantum field mechanics for one elementary field-particle which preserves especially the property of physical unity and the ensuing fundamental coherence. We have always one and the same 'world sandwich' prepared from two main protofields, the dynamically perceivable electromagnetic proto-medium (alias *ether* corresponding to the 'electromagnetic potential' and remaining *both* real and unobservable in its unperturbed, 'pre-Creational' state), in which we actually 'live', and the directly inaccessible (and therefore not well specified by its exact nature, but nevertheless quite real) gravitational background which manifests itself only through its coupling to the electromagnetic side of the 'sandwich' (although the latter appears as our three-dimensional world, it is initially created in the 'embedding' space of still more fundamental level of complexity currently inaccessible for us and therefore 'incomprehensible'; it is difficult to state something about its 'dimensions'). The coupling interaction between the protofields gives rise to the complex, or 'quantum', behaviour of the elementary field-particles according to the above FMDF picture. There seems to be two main species (or regimes) of coupling corresponding to only 'electro-weak' forces (leptons) and 'strong' (including electro-weak) forces (hadrons), with many individual field-particles within each species (other types of coupling cannot be excluded). Those fundamental interaction species can be considered themselves as a result of the two-fold (in general, *n*-fold) dynamic redundance (splitting) of a single basic process of interaction (coupling) between the two component protofields. The well-known existence of several similar 'generations' of the fundamental 'bricks' of the world, remaining basically unexplained and looking as if 'excessive' within the unitary-science logic, is also consistently inserted into the new world picture as a result of the same dynamic redundance of the fundamental interaction process which never cannot, and practically should not, be limited to a 'strictly necessary', unrealistic minimum. The same profound consistency with the dynamic redundance paradigm involves the observed multiple unstable species of 'elementary' particles and 'resonances', with their characteristically 'broken' symmetries (the universal symmetry of complexity is *always* naturally 'broken', but contrary to the linear-science concept of broken symmetry, it is a much *higher* symmetry than any mechanistically exact, 'unbroken' prototype, see also below). Many of the observed 'elementary' particles including the 'exchange particles', such as the photon, are not independent objects as it is explained above and move at the electromagnetic 'side' of the world, whereas the existence and observed manifestations of similar 'exchange particles' at the inaccessible gravitational side and the corresponding 'gravitational waves' seem to be more ambiguous and need further investigation. A considerable modification is needed also for the linear-science concept of permanent creation and destruction of 'virtual' (including *massive*) particles simply 'in vacuum' (the 'polarisation of vacuum'), which in its existing version directly violates the complexity conservation law (contrary to the linear-science version of 'energy conservation' which can formally be violated, if 'it is only for a short time'). The physical meaning of 'fermionic' and 'bosonic' species as well as some basic particular cases of the many-particle behaviour, like e. g. Bose condensation, are also causally clarified within the same complex-dynamical picture.

In summary, we emphasize once more that *one and the same* complex-dynamical, *physically based* process of quantum beat gives *all* the known fundamental manifestations of the unreduced complexity always remaining unexplained and 'mysterious' within the canonical, unitary science: (i) quantum wave mechanics of the micro-objects (wave-particle duality and fundamental quantum indeterminacy) associated to causal origin of space and time; (ii) 'special relativity' (time and space involvement with motion) associated to causal origin of electric charge and spin; and (iii) universal gravitation (universal gravitational attraction of all massive bodies, time and space involvement with mass and gravity, and fundamental discrete structure of space and time, or 'quantum gravity') associated to the causal, universal origin of mass.

We can clearly see now that the outline of the 'brief history of time' (Hawking 1988), as well as other similar compilations of the fundamentals of the unitary physics, massively commercialised by the single-valued, linear science is indeed *too* brief and especially over-simplified with respect to the underlying 'Weltanschauung' pretensions and the intrinsically 'living', self-developing, irregular and unlimited diversity of the holistic *real* world around and within us. The self-chosen, arrogantly pretentious 'sages' evidently cannot even propose any clear understanding of the causal, objectively transparent nature of the very fundamental entities of time, space, elementary particle, its intrinsic duality, mass, charge, spin, other basic properties, but they do not hesitate to assert not only that they *understand* how Nature is, but also how it was created and evolved to the present state, basing their conclusions on the obviously contradictory and incomplete correlations within the purely abstract, arbitrary systems of symbols, where everything can be 'deduced' and 'justified'. Actually it is that kind of mechanistic, trickily manipulated 'history' which is certainly finished now, in both the knowledge and society development (see also Prologue). The extended causality of the universal science of complexity shows unambiguously that the canonical science has been nothing but an extremely restricted, probably inevitable, but definitely condemned to failure, attempt to understand the genuine richness of the developing hierarchy of complexity, and now the ending 'brief', reduced history of time and humanity should be definitely replaced by the fully conscious, causally complete version capable to provide the genuine, intrinsically creative understanding of the developing reality. The esoteric 'pulp fiction' from science, promoted by the exhausted medieval mechanism, with all its artificial, ambiguous sophistications and over-emphasised, superficial 'weirdness' basically detached from reality and thus accepting any formal 'possibility' in a subjective 'competition of influences' should give place to the reality-based, objective transparency of the universal complexity unfolding process.

The chaotic dynamics of quantum objects at the next higher level of complexity, naturally emerging from that of the elementary field-particles, corresponds to either *quantum measurement* process (partially dissipative dynamics of an open system) or Hamiltonian *quantum chaos* (conservative dynamics, closed system), both presented now in their complex-dynamical, causally complete versions and described by essentially the same universal equations and method of their analysis as the fundamental reduction-extension dynamics at the first level of complexity. This universality of description reflects the basic unity of the real world it represents and the ensuing similarity of the physically occurring complex-dynamical processes at any level.

Thus, quantum measurement is the physically real process of the field reduction (self-sustained squeeze) induced by any 'ordinary', higher-level interaction (typically of electromagnetic origin) with the real excitation of other degrees of freedom (levels of complexity) and governed by the same mechanism as the quantum-beat reduction at the first level. It can also be considered as a result of a *transient* 'bound state' of two or more participating elementary field-particles which continue their respective quantum-beat oscillations 'inside' the 'embracing' quantum measurement process, where the corresponding 'extended' phases of the elementary fields are actually localised in a close vicinity of the common 'centre of measurement (reduction)', which suppresses the manifestations of their 'wave' properties, like diffractive interference. This rigorously substantiated picture of the realistic 'wavefunction collapse' explains the observed 'peculiar' properties of quantum measurement postulated in the conventional quantum mechanics, like the apparently 'classical' behaviour of the 'measuring instrument' (the system behaviour becomes indeed localised and therefore classical, but only *transiently*, see also below for the general interpretation of the classical type of behaviour).

The Hamiltonian quantum chaos is based on the same standard complex-dynamical sequence of reductions and extensions, but it happens in the absence of any noticeable real excitations at other levels of complexity and therefore instead of spatial squeeze involves rather a momentum- or configuration-space reductions to particular realisations of the effective interaction potential forming the permanent process of chaotic realisation change (one can also have a combination of quantum chaos and quantum measurement regimes). It is clear that we deal here with the *true* quantum chaos possessing purely dynamical randomness and not only some particular form of regular (unitary, single-valued) dynamics endowed with 'signatures of chaos' or a capacity to 'amplify environmental influences' always depending on the time of observation and other parameters as it happens in multiple linear-science simulations of quantum chaos, or 'chaology' (Eckhardt 1988, Gutzwiller 1990, Berry 1991, Bohigas 1991, Haake 1991, Bogomolny 1992, Prigogine 1992, Ikeda 1994, Izrailev 1995, Chirikov 1995a,b, Zhang and Feng 1995, Andreev *et al.* 1996, Brun, Percival, and Schack 1996, Casati 1996, Prosen 1996, Shigehara and Cheon 1997, Bogomolny *et al.* 1997, Zurek 1998). Application of our analysis to a generic case of Hamiltonian quantum dynamics (Schroedinger equation with periodic potential) shows how our version of quantum chaos naturally passes to its classical analogue (also eventually extended from its linear-science simulation) under the ordinary *quasiclassical transition*, which reconstitutes the *principle of correspondence* in quantum mechanics directly violated in the mechanistic simulations of quantum chaos (cf. Ford and Mantica (1992)). The truly unpredictable quantum dynamics of Hamiltonian systems is obtained in terms of the Schroedinger equation and wavefunction, rather than density matrix in which the additional randomness, or 'decoherence', is actually axiomatically *inserted* by the definition. This explains why the linear-science studies of quantum dynamical randomness use so often the density-matrix or similar formulations and why the results are basically deficient: they will inevitably contain a logical vicious circle, in one or another form. In contrast to this, in our approach we rigorously *derive* the extended version of the density-matrix type of description revealing the causal origin and physical meaning of the additional randomness within a 'mixed' state. This additional unpredictability is a higher-level manifestation of the *same* universal origin of dynamic uncertainty (complexity) that gives fundamental quantum indeterminacy at the lowest level, described above.

Further increase of dynamic complexity leads to its next higher level represented by the *classical type of behaviour* and starting approximately from the most elementary *bound systems* (like atoms) which naturally manifest the characteristic property of classical trajectory localisation. The latter is a result of the specific, causally random character of the underlying quantum-beat dynamics: since the corpuscular state of *each* of the bound fields 'chooses' the position of the next centre of reduction in a causally random, independent fashion (though usually not without an *average* tendency in the *probability distribution*), while the *bound* fields tend to remain very close to one another, they can perform the absolute majority of their cycles of reduction-extension only in the close vicinity of their common 'centre of mass'. The 'correlated' jumps of the bound field-particles to larger distances are not impossible, but they will normally be very rare, unless they are stimulated by a well specified additional interaction with other bound systems (as it happens in all 'macroscopically coherent' states with 'Bose condensates'). The essential condition of independent chaotic wandering of each elementary virtual soliton is defined by the fact that the fundamental electro-gravitational coupling within each particle is much stronger than its binding to the interaction partners, which is a natural demand for an individually occurring elementary particle (see also below, the discussion of the modified Planckian units). Note the clear distinction of our definition of the classical, quasi-localised, 'trajectorial' type of behaviour, related to the appearance of a next higher level of complexity, from the canonical 'semiclassical (quasiclassical) case' corresponding simply to the relatively small value of the de Broglie wavelength and only externally simulating certain observable properties of the truly classical behaviour (cf. e. g. Holland (1995)). Although both cases normally correspond to increased dynamic complexity with respect to a typical 'wave' behaviour and thus generally 'go in the same direction', the emergence of a truly classical regime needs a specific qualitative change of a 'phase transition' to a higher level of complexity, actually taking the form of the bound-system formation. The 'decoherence' due to some 'external' influences can contribute to the degradation of the characteristic wave properties of a system, but it definitely *cannot* constitute the *fundamental* origin or character of the classical behaviour, contrary to various recent approaches within the single-valued paradigm (e. g. Paz and Zurek (1993), Schack, d'Ariano, and Caves (1994), Omnès (1995), Shiokawa and Hu (1995), Griffiths (1996), Zeh (1996), Anglin and Zurek (1996), Zurek (1997), (1998)). The dynamic redundance paradigm provides a much better, intrinsically dynamical, always present source of internal 'decoherence' (which is causally specified as *incompatibility* of redundant states, as opposed to the linear-science 'loss of coherence' of *superposable* states), so that now one does not need to rely upon always changing, ambiguous 'environmental influences' while trying to explain an evidently universal, fundamentally rooted phenomenon. In terms of formal description, this means that various linear-science versions of the 'density matrix formalism' or 'consistent (decoherent) histories' *based* on the explicit *postulation* of the statistical (randomised) behaviour and often used in decoherence type of simulations of quantum indeterminacy and reduction possess the evident logical inconsistency already mentioned above. Contrary to this, in our approach we rigorously *deduce* (and therefore causally *define*) randomness within a purely dynamical, non-statistical description and thus obtain a *causally* statistical description of the next level (the 'generalised Schroedinger equation', see also below) which considerably extends the existing simulations of the density-matrix type.

     Note in this relation the same kind of deficiency within a broader scale of similar attempts to 'clarify' the canonical quantum mechanics which suffer from the same characteristic, fundamentally inevitable 'impotence' of the mechanistic approach when in order to obtain an effect one should first explicitly insert it 'by hands' in one form and then develop the 'consequences' in another, trivially equivalent form  (i. e. from the beginning *tout est donné*, see Bergson (1907)) creating thus the evident 'vicious circle' à la *deus ex machina* (e. g. Ghirardi, Rimini, and Weber (1986), Pearl (1989), Ghirardi, Grassi, and Pearl (1990), Ghirardi, Grassi, and Rimini (1990), Gisin and Percival (1992), (1993a,b), Diósi (1992), Dove and Squires (1995), Brun (1995), Brun, Percival, and Schack (1996), Brun, Gisin, O'Mahony, and Rigo (1996), Brun (1997), Kist *et al.* (1998), Goldstein (1998), Kent (1998)). It is not surprising that none of those approaches can ever really pass to a realistic, physically based description and they are obliged instead to operate with 'vectors' and their 'projections' in abstract 'spaces', 'matrix elements' and other formal constructions *irreducibly* detached from reality. By its purely speculative, non-creative character, this kind of 'new physics' is close to another group of simulation reduced to infinite reformulation of 'quantum enigma' and concentrated especially around the related properties of 'quantum nonlocality' and 'indeterminacy'. The formally defined 'consistent histories' and other speculative 'elements of reality'; purely formal 'inequalities' and the related fruitless experimentation around 'nonlocality'; the never-ending series of speculative 'quantum-mechanical' games and 'gedanken experiments' with abstract 'creatures' and paradoxes; fruitless, purely linguistic or philosophical 'interpretations' and 're-interpretations' around the same, unchanged reality and its unresolved 'mysteries'; evidently figurative and unrealistic, but excessively 'promoted' and 'seriously' discussed quantum 'clonings', 'teleportations', 'unitary computations', 'time machines', 'tomographies' and 'endoscopies'; macroscopic 'electromagnetic coherence' in living organisms and 'decoherence' of 'histories' and other formal constructions; 'quantum logic' and 'quantum algebra' - the enormous flood of all these and many other similar 'developments' of the last period shows that the phenomenon of 'impostures intellectuelles' has had a much wider extension than it was originally noted (Sokal and Bricmont 1997), and is actually transformed now to the dominating pseudo-scientific parasitism, actively stopping any truly progressive, realistic development of knowledge. It is easy to see that all those branches of 'quantum mysteriology' add absolutely nothing to the essential canonical knowledge as it is expressed by the empirically based postulates of the standard interpretation of quantum mechanics, honestly recognising the objective border between the known and the unknown, while the vain sophistication of the 'mathematical quantology' is reduced to tricky, but evidently fruitless 'replays' of the same situation in the form of ever more odious mystification of abstract symbols and superficially esoteric 'mind games'. Despite their oligarchic dominance in the bureaucratised scientific structures and media, the adherents of the 'advanced symbolism' can never reveal what the *exact, physical nature* of the existing entities is and what *really happens* to them. Instead, the intrinsic *imitators of creation* silently 'borrow' the features 'to be expected' from the suppressed realistic approaches and arrange for their superficial fitting into the promoted mechanistic constructions, which can always be achieved in the linear science by playing with parameters, logical rules, etc. They tend to call it 'explanation' and 'understanding'. In the evident absence of any real solution, the proliferating mediocrity concentrates its 'intellectual' activity around all kind of 'informally' hidden, oligarchic 'self-organisation' at all levels in order to impose its formal, mechanistic dominance which parasitically abuses formal 'neutrality' of the 'developed' Unitary System of organisation and results in complete blocking of the really free, progressive development of the fundamental knowledge just at the time when it is most needed. The 'réel voilé' (d'Espagnat 1994) is thus transformed into the 'réel violé' by the tricky adherents of the self-'chosen' community of modern 'scribes and Pharisees' accepting any degree and kind of deviation from the elementary causality and consistency in order to promote their unjust, oligarchic dominance in scientific institutions and sources of information and financial support. Being intrinsically unified with the similar directions of 'chaology', 'complexology' and 'mathematical (symbolical) physics', this anti-constructive noise is accompanied by unprecedented publicity campaigns strikingly resembling those in

massive commercial promotion of second-rate products and finds a 'strangely' generous support from the 'friendly' bureaucracy at the time of general financial 'difficulties' in science and thus considerably, and deliberately, reduces the chances of renaissance of the genuine, intrinsically realistic creation that can only be based on the *explicit, complete, physically based solutions* supported by a transparent, non-contradictory system of correlations within the whole knowledge (see also below).

A number of particular, practically important consequences is obtained and outlined within the new understanding of the fundamental reality introduced by the quantum field mechanics. They involve the values of the Planckian units and the causally substantiated structure of the expected spectrum of elementary particles; quantum stages of the black hole evolution; impossibility of creation of the unitary 'quantum computer' (simulator), proposed within the linear approach, and possibility of creation of a 'chaotic (complex-dynamical) quantum simulator'.

Using the above unified picture of the quantum field mechanics, one can easily show that the well-known Planckian units of the smallest existing scales of space and time and the largest value of elementary-particle mass should be renormalised, which permits one to reconstitute the realistic, and practically meaningful, values of the extreme quantities. Indeed, the above causal interpretation of gravity as an *indirect* result of the complex-dynamical electro-gravitational interaction process shows that the measured effective magnitude of the force of that indirect gravitational attraction between the bodies, characterised by the usual 'gravitational constant' which is used for obtaining the conventional values of the Planckian units, need *not* coincide with the magnitude of the *direct* electro-gravitational coupling which is *actually* responsible for the smallest attainable spatial dimension, the size of the squeezed state of the elementary field-particle. Therefore we should substitute a new, rescaled value of the 'gravitational constant' into the expressions for the Planckian units, the one that characterises the direct fundamental interaction between the electromagnetic and gravitational protofields. It is clear that it should be much higher than the ordinary value, so that the integral internal dynamics of an elementary particle, for example the electron, could be maintained despite its various external interactions with other particles (in particular, the e/m interaction between charged particles, known to be much stronger than their observed, indirect gravitational interaction, is at the same time much weaker than the direct coupling of each of them to the gravitational medium). In order to specify the magnitude of the necessary rescaling, we suppose that the new value of the Planckian unit of length coincides approximately with the lowest length scale appearing experimentally, i. e. $10^{-17}$ - $10^{-16}$ cm, which shows that the ordinary value of the gravitational constant in the corresponding expressions should be multiplied by $10^{33}$ - $10^{34}$. Then we can *verify* the assumption made by calculating the new values of the Planckian units of time and mass and comparing them with the known extreme values of time and mass. We obtain indeed a good agreement, with the new Planckian value of mass of the order of 100 GeV ($10^{-22}$ g) and the Planckian time of the order of $10^{-27}$ - $10^{-26}$ s. This result resolves a number of paradoxes with the excessively extreme ordinary values of the Planckian units (in particular that of the unrealistically big unit of mass creating a huge 'empty space' in the expected mass spectrum of particles), and leads to the very important *practical* conclusion that there is no sense to look for new particles at the space scales much smaller than $10^{-17}$ cm and mass scales much greater than 100 GeV, which allows us to avoid practically difficult further increase of the accelerated particle energy (the values of the extreme quantities can be subjected to a reasonable further refinement which will not change the principle). This conclusion is supported also by the 'self-sufficient' general picture of the quantum field mechanics demonstrating a quite complete, harmonious, and realistic construction of the world at its fundamental levels. What we need is further clarification of the details of that *causally complete* general picture *within* the realistic scales designated above (in particular, the appearing experimental indications on the existence of very small objects within the known truly elementary particles can be consistently interpreted as being produced by the squeezed states of the respective quantum-beat processes possessing indeed the minimum observable sizes). The problems involved can be efficiently resolved only with the help of the *physically* based picture of the unreduced, complex dynamics of the elementary objects providing the unique reduction of the large number of existing formal possibilities, in accord with the natural complexity development process and contrary to the basically ambiguous, purely symbolic sophistication of the canonical, linear-science approach.

Another practically important conclusion concerns the actively developed proposition to create the so-called 'quantum computers' whose operation, and the expected advantages, are based on the unitary (single-valued), and thus non-dissipative, regular and reversible, evolution of the canonical quantum mechanics (Benioff 1980, 1986, 1997, Feynman 1982, 1986, DiVincenzo 1995, Ekert and Jozsa 1996, Lloyd 1996, Steane 1997, Boghosian and Taylor 1997). The *fundamental* impossibility of realisation of such a device directly and unambiguously follows already from the actual *absence of the unitary evolution* at any level of the *real* world dynamics explicitly revealed in the quantum field mechanics, since otherwise it would be a single-valued, effectively one-dimensional, unrealistically predictable (Laplacean) world. Specifically, any nontrivial interaction in course of computation, and in particular irreducibly irreversible function of any *memorisation*, will lead to the fundamental complex-dynamical redundance and instability with the unavoidable failure of any single-valued (unitary) scheme. Unitarity of a computer would necessarily mean the integrability of its dynamics, according to the causally clarified concept of integrability of the universal science of complexity, and this implies that at best it would be capable to 'numerically solve' only integrable, analytically resolvable problems. This is a consequence of the general 'complexity correspondence rule' following from the complexity conservation law and stating that a computer can correctly simulate only problems with dynamic complexity not exceeding its own complexity (including eventually interaction with the complex 'environment'). In terms of 'quantum bits' so popular among the unitary 'quantum programmers' we can say also that each 'bit' (and not only a quantum one) corresponds to a complex-dynamical realisation and can physically be produced or registered only in the irreversible, partially irregular fashion involving complex-dynamical, non-separable (fractal), irreducibly probabilistic entanglement of the interacting entities. Whereas the higher-level, classical bit chaoticity can be sufficiently (though never totally) 'controlled', any attempt of control of chaos of a bit-realisation at the most fundamental, quantum level evidently either produces no effect at all, or introduces still greater, 100 per cent uncertainty in the bit dynamics. The impossibility of creation of regular (unitary) quantum computers and simulators is compensated by the indeed practically interesting possibility of creation of *chaotic (complex-dynamical) quantum simulators* which can open new prospects for the efficient real (complex) problem solution, but need the unreduced, dynamically multivalued analysis of the universal science of complexity.

The hierarchy of levels of complexity develops in the same fashion to the more and more involved systems always preserving, however, the universality of the complexity origin and its main properties, as well as the intrinsic dynamic continuity between the levels equivalent to the *universal correspondence principle*. All the universal manifestations of complexity, such as space, time, and energy, have the same hierarchical structure with fractally branching, causally probabilistic levels-realisations.

It is shown that such complete universality of the dynamic complexity concept can be expressed in a more explicit way by a standard formalism of description of arbitrary system behaviour at any level of complexity. It is based on the most fundamental principle of nature, the *complexity conservation law*, or *universal symmetry of complexity*. It states that the dynamic complexity universally determined by the number of system realisations (or, equivalently, as the rate of their change) cannot appear or disappear in an arbitrary fashion, but is always conserved for the closed systems, or can be transformed in a well-specified fashion in interaction processes involving open systems. As the complexity of the closed systems is conserved, it inevitably undergoes permanent *internal transformation* from a 'hidden' (latent) form of interaction, or potential energy, universally referred to as *(dynamic) information* into the second, explicit form of spatial structure, universally defined as *(dynamic) entropy*. The transformation process is 'spontaneous' in the sense that it is driven by the interaction itself that represents therefore intrinsically unstable form of complexity in which it exists immediately after its creation (emergence) in the form of a new system (level) of interacting objects. This universal, eventually hierarchical, transformation of the initial stock of informational complexity, encoded in *real* though latent interactions, into the incarnate, tangible form of spatial structure-entropy is the basis for all dynamical processes in nature, and thus for any existence. Correspondingly, it can be described by the universal dynamic equation, generalising and extending all the known equations that correctly describe real processes. This is the *(generalised) Hamilton-Jacobi equation* for the action-complexity (which considerably extends the notion of the ordinary mechanical action) having the form of its well-known analogue from classical mechanics, but possessing much wider interpretation and unlimited scope of applications. We prefer to call this unified description *Lagrange-Hamilton equation* (formalism) taking into account the close involvement of the Lagrangian approach and its *universal*, and so much missing until now, fusion with the Hamiltonian method. The unified equation reproduces all the known equations of science as its particular limiting cases and approximations and also proposes a new approach to their solution and analysis coinciding with the universal formalism of the fundamental dynamic redundance (the method of the effective dynamical functions). This unrestricted universality is based on the universal character of the underlying notion of complexity and its conservation law that includes, in a considerably extended form, such linear-science principles as energy conservation (and all other well-established physical conservation laws), the second law of thermodynamics ('degradation of energy' through permanent increase of entropy), the 'principle of relativity' (special and general), and various versions of the 'variational principle' ('least action', 'shortest optical path', 'minimum potential energy', etc.).

Note that the generalised, dynamic entropy can also be presented as the complexity of the generalised spatial structure, both of them *appearing* at a certain level of complexity, and therefore being *physically real*. In a similar fashion, time is naturally associated to the *disappearing* form of the same complexity, the dynamic information, and therefore time is real and perceivable only as the sequence of events in the *same* process of complexity transformation, but *not* as a physical, material entity, or 'dimension'. Since the universal Lagrange-Hamilton equation describes the complexity transformation, it actually takes the form of relation between the temporal rate of disappearance of potentially real (informational) possibilities and spatial rate of appearance of materially real (spatial) structures within the *same* process, which is none other than the generalised *dispersion relation* (between the generalised energy and momentum or, equivalently, frequency and wave vector).

At each level of complexity (or for each complex system) the extended, dynamically multivalued description of the science of complexity provides the complete (general) solution of the universal equation in the form of many incompatible realisations replacing the single realisation of the linear science. *All* the diverse realisations of a system are profoundly *symmetric* among them, in the extended sense of the universal dynamic symmetry of complexity and irrespective of, or even *due to*, their *mechanistic*, formal configurational asymmetry. This property of the symmetry of complexity is simulated by the linear-science notion of 'broken symmetry' trying to account for the empirically observed large deviations from the simplified linear-science symmetry. However, in accord with its name any 'broken symmetry' is a *lower* symmetry with respect to the corresponding 'unbroken' mechanistic symmetry, while the unreduced dynamic symmetry of complexity between the formally 'asymmetric' realisations is qualitatively *higher* than any its mechanistic prototype. Needless to say, the higher is the system complexity, the higher is the corresponding dynamical symmetry, so that any formally 'irregularly structured' living organism is indeed much more symmetric than any most involved, or simple, regular structure. This fundamental conclusion has been previously expected, but it acquires a rigorous, causal interpretation and substantiation only now, within the dynamic redundance paradigm.

The complete solution of a version of the universal equation realises the fully adequate description of all the involved complexity patterns for each level of complexity, system or type of behaviour. They are but particular manifestations of the universal symmetry (conservation) of complexity in the form of the standard process of *complexity unfolding* (development) where the full *positive* complexity of a system is first autonomously created from lower-level (more fundamental) structures in the predominantly informational form of a potential (latent) interaction specifying the notion of *élan vital* (Bergson 1907; among many other similar and related ideas like Aristotle's *entelechy*, or *vis vitalis* from various vitalistic concepts) which is then transformed into the form of (generalised) dynamical entropy of real, tangible structures in the process of interaction development by dynamical realisation breeding, so that the sum of dynamical information and entropy remains constant and equal to the total dynamic complexity of the (isolated) system. This shows that not only the canonical postulated 'second law of thermodynamics' inexplicably and unpleasantly 'skewed' towards 'degradation of energy' is extended now to the causally deduced and *dynamically* substantiated universal *symmetry* (*conservation* principle) of dynamic complexity, but also the processes of *both* degradation *and* emergence (development) of any structures are none other than different, objectively inevitable stages of the *same unique process* of now rigorously and causally defined *birth, life and death* of *any* dynamical system. The *apparently* 'progressive development' of a system is evidently closer to its birth from lower-level structures, while the phase of *explicit* 'degradation' is a precursor of its death, or generalised *state of equilibrium*, thus also acquiring a causally extended, universally substantiated interpretation. However, the system degrades, strictly speaking, during its whole life, i. e. starting already from its birth, while the life of even very *old* system is irreducibly *based* on the continuing internal *progressive* transformation of the remaining (small) stock of information into entropy. This universal interpretation of development is related to the fact that the natural emergence of form as a regular *tendency* in the distribution of realisation *probabilities* always corresponds to *growth* of the unreduced dynamic entropy-complexity (at the expense of the diminishing information), whereas in the canonical science any ordering is associated to a *decrease* of entropy, and therefore the 'spontaneous' structure formation is incompatible with the 'second law' postulating permanent entropy growth.

Since the structure of complexity is dynamically *quantized* (discrete), the universal information transformation into entropy (or system life) consists of a number of more abrupt 'jumps' of development (or degradation) alternating with periods of more even change (this is the generalised 'punctuated equilibrium'). The steady change periods correspond to the internal development of a level of complexity-entropy forming a tendency (structure), which is the complex-dynamical extension of the 'second-order *phase transition*' of the disorder-order type (starting from the 'disordered phase' at the beginning of a level of entropy). Similarly, the larger 'jumps' between (big) levels of complexity extend the notion of the 'first-order phase transition', where the new level of complexity-entropy corresponds to the new 'phase' in the conventional description mechanistically imitating only one, limiting regime of 'uniform chaos'. The fractal structure of complexity reproduces itself in the general hierarchy of phase transitions, where all the intermediate 'orders of transition' can occur, as well as in the fractal-like structure of a given transition containing a variety of 'precursors' and 'germs of the new phase'.

The self-consistent source of intrinsic, autonomously maintained *creativity* of the *natural* evolution of a system through its complexity development is thus revealed as an extension and substantiation of the famous Bergsonian *évolution créatrice* (Bergson 1907). The probabilistic, fractal structures emerging at a particular level of complexity unfolding give rise to the next generation of new objects *together with their interaction* realising further development of the dynamical creation process described again by the universal Lagrange-Hamilton formalism. This rigorously specified and universally consistent picture of the *creative evolution* of a system is based on a crucial extension of the notions of information and entropy, so limited and very often misunderstood and confused in their conventional, purely formal and characteristically vulgarised, versions in the linear science (e. g. Brillouin (1956), Nikolis (1986), Haken (1988), Žurek (1989), Kadomtsev (1994), Cerf and Adami (1996), Izakov (1997), Van Siclen (1997), Lyre (1997), Steane (1997)). Indeed, the single-valued paradigm of the canonical, mechanistic science can never provide any intrinsic, dynamical source of multiple independent, probabilistically realised 'states' indispensable for any consistent definition of information and entropy. The universal symmetry of complexity shows also that *both* energy and entropy are different expressions of the same dynamic complexity, where the (dynamic) entropy is more directly determined by the number of realisations for a fully developed system or level of complexity, whereas energy characterises the temporal rate of realisation change in the process of complexity unfolding, usually at a lower, relatively fine-grained level of complexity (similarly, the generalised *momentum* characterises the spatial rate of realisation emergence in the complexity transformation process). It becomes clear now that the canonical thermodynamics and relativity as well as other branches of the 'classical' physics operating with the notions of entropy and energy deal actually with a particular limiting regime of complex dynamics, the 'uniform chaos', represented by sufficiently frequent chaotic transitions between closely spaced realisations. This chaotic 'internal structure' of the externally uniform dynamics is not properly recognised in the reduced notions of the linear science, which gives characteristically 'irresolvable' problems like that of irreversible 'degradation' of energy ('arrow of time').

The generalised Lagrange-Hamilton formalism possesses two universal, equally valid, and intrinsically related forms providing respectively the *local* (trajectorial) or *nonlocal* ($\psi$-functional) type of description. The local type of description corresponds to system occupying 'normal' realisations one after another – which gives the generalised system *trajectory* in the generalised space (of realisations) at the corresponding level of the hierarchy of complexity – and is represented by the Lagrange-Hamilton formalism, but especially its (generalised) *Lagrangian* version that describes, in the science of complexity, all possible trajectories-realisations *really*, though probabilistically, taken by the system in course of its *chaotic transitions from one trajectory (realisation) to another* driven by the fundamental mechanism of dynamic redundance. This causally deduced description not only provides the realistic refinement of the least action conjecture in classical mechanics (and similar 'variational principles' in other fields of the canonical science), but universally extends it to any dynamical system behaviour upon which it becomes indistinguishable from the *modified* path-integral formulation of a problem (the regular, effectively one-dimensional systems with only one continuous trajectory-realisation form a set of trivial and rare exceptions with the zero relative measure). The nonlocal version of the same universal formalism corresponds to the state of the system just undergoing those chaotic transitions between realisations (elementary trajectory portions) and is represented by the *generalised Schroedinger equation* for the causally extended $\psi$-function describing that particular transient state of the system *between* the realisations. According to this definition, the $\psi$-function gives the (dynamically determined) *distribution of realisation probabilities* (i. e. probabilities of realisation occupation by the system), and therefore plays the role (for certain levels of complexity in the form of its squared modulus) of the (generalised) *distribution function*. The obtained Lagrange (local) and Schroedinger (nonlocal) descriptions represent two intrinsically (dually) related aspects of the same dynamical process of complexity unfolding (alias 'system evolution') and therefore are related by the causally derived generalised 'quantization rules'. At the level of quantum-mechanical objects the generalised $\psi$-function takes the form of the canonical *wavefunction*, now acquiring the transparent causal interpretation of a *physically real entity* (see above). For the classical systems, at higher levels of complexity, the same universal $\psi$-function turns into the corresponding (modified) distribution functions describing now causally probabilistic, *dynamically* uncertain distribution of the quasi-localised state-trajectories (and the system performing chaotic transitions, or 'quantum jumps', between them). Therefore the $\psi$-functional representation of complex dynamics can also be considered as the unified causal extension of the path-integral and variational formulations of the linear science (where the least action principle turns out to be *more* universal than the path-integral formalism, so that in the generalised version the latter is reduced to the former).

In general, the system evolution can be considered as the chaotic piece-wise total trajectory consisting of more uniform sequences of (probabilistically taken) realisations interrupted by probabilistically distributed larger 'jumps' between such quasi-uniform pieces of trajectory. If the driving interactions involved are such that irregularity is relatively high and the quasi-uniform pieces of trajectory are relatively short (in the limit each of them includes in average only one 'point'-realisation), then we deal with a pronouncedly nonlocal (distributed), 'wave-like' behaviour more naturally related to the nonlocal version of the universal formalism. In the other limiting situation the irregularity is relatively low, the chaotic jumps between realisations are typically small, and we obtain relatively long quasi-uniform pieces of generalised trajectory (in the limit it is transformed in the single quasi-continuous Newtonian type of trajectory of a well-defined 'material point'), in better correspondence with the local, properly 'trajectorial' version of description. However, both dual pictures, aspects of behaviour and versions of its description are always present and applicable.

We clearly see now that higher levels of complexity acquire many properties of causally interpreted 'quantum' and 'relativistic' behaviour, whereas quantum systems can be naturally described in terms of the modified 'classical' (real) field and particle mechanics (in the latter case the trajectorial type of description provides the causal, Lagrangian extension of the path-integral formulation of quantum mechanics). Phenomena and properties like 'wave-particle duality', 'quantization', 'quantum jumps', 'production (creation) and destruction (annihilation) of particles/realisations' ('second quantization'), uncertainty relations, 'quantum tunneling', 'special and general relativity', etc. not only obtain their realistic, complete interpretation for elementary particles (fields), but are rigorously reproduced for *arbitrary* complex dynamics and thus practically for *any kind of dynamical system*. Thus the universal property of *dynamical discreteness* (intrinsically quantized character) of system behaviour can be understood as a direct consequence of its *globality* fixing the relevant quantum of complexity and resulting in its turn from the natural property of *integrity* of a system implying that within it 'everything interacts with everything' and therefore arbitrary (small) changes are impossible. The same universality refers to the generalised 'principle of complementarity' which can be considered now as the summarised expression of the various characteristic dualities (self-sustained intrinsic alternatives) of arbitrary complex dynamics directly related to the fundamental dynamic redundance paradigm and therefore inexplicable within the unitary science (thus the above-mentioned discreteness of system realisations is the dual complement to the intrinsic continuity, or integrity, of its dynamics). The 'quantum (wave) mechanics', in the narrow sense, is seen now as several lowest observable levels of the dynamic complexity of the world, possessing already all its universal properties which appear in their simplest, 'canonical' form and create the notorious 'quantum paradoxes' unresolvable within the single-valued, unitary science and therefore providing an ideal subject for 'quantum mysteriology' (in a similar fashion, various branches of the hopelessly single-valued 'complexology' and 'chaology' desperately tackle the multivalued reality at higher levels of complexity, cf. Horgan (1995)).

Within the same analysis of arbitrary complex behaviour we propose also universal extensions of other known linear-science approaches to complex dynamics, such as the geometrical analysis of the *Kolmogorov-Arnold-Moser (KAM) theory* and the concept of *Poincaré surface of section*.

The appearance of a next higher level of complexity from the interacting entities of the preceding lower level can be summarised as the *dynamic entanglement* of the degrees of freedom of the interacting entities, necessarily accompanied by the dynamic redundance of the number of entangled combinations. The dynamic entanglement is the profound, inseparable, fractally structured and causally probabilistic mixture of the interacting degrees of freedom, which considerably extends the linear-science versions of mixing and entanglement, used in particular in quantum mechanics, as well as that of arbitrary 'interaction process' which is also nothing else than the dynamic entanglement (it is clear therefore that the latter is another expression of the universal dynamic fractality described above). The irreducible redundance of the products of an elementary interaction process is due to the fact that both before and after the interaction process we have the *same unique reality* with a fixed number of 'places' for the degrees of freedom. However, if each of the interacting entities is characterised by $N$ degrees of freedom, their combination in course of dynamic entanglement-interaction will produce the equivalent of $N^2$ versions of the resulting combined degrees of freedom (characterising the new-born entities). The underlying 'conservation of the total number of degrees of freedom' is none other than a manifestation of the complexity conservation law. We obtain thus the $N$-fold redundance of the created structure at the appeared new (higher) level of complexity. That is how inseparability (real and mathematical) of the dynamic entanglement (mixing) is related to the dynamic redundance. It is important that the redundance (local multivaluedness of reality) exists in a *real* form even *before* its actual emergence in the process of interaction-entanglement, it is hidden in the 'latent' form of interaction-potential (it is the generalised 'energy of wound-up spring'). When the redundant realisations *actually* appear in the process of interaction, they start inevitably replacing one another by pushing each other out from the 'tangible', explicit reality back into the hidden form of interaction which is, however, not stable any more, and thus the chaotic 'quarrel' of the competing realisations continues, until the appearing next level of interaction splits and entangles them into a still higher level of the fractal structure of complexity, etc. (this 'complexity unfolding' continues till the complete exhaustion of the total, initially hidden, complexity of the global isolated system). At each level of complex dynamics the redundant, multi-faced result of interaction preserves, however, its wholeness, which is described just by the generalised $\psi$-function (distribution function) consisting of all the (real) transient states of the compound (mixed) system during its chaotic jumps between each two successive realisations and playing the role of the universal dynamical *'glue'* that bonds all the multiple realisations of the complex dynamics into a single system.

When the system is in this transient 'phase of $\psi$-function', it 'transforms' its structure from the last to the next realisation and is therefore 'obliged' to pass by a loose, transiently *disentangled* state of interaction partners (entities) temporarily *returning* to their initial, quasi-free state, similar to the particular case of the elementary field-particle dynamics described above (cf. also the 'asymptotic freedom' of quarks, which now seems to be of the same *dynamic* origin, whereas the quarks themselves are but unstable interacting entities/realisations from a lower level of complexity). This shows also that *each* interaction process, irrespective of its external appearance, has its internal 'pulsating' structure determined by the cycles of reduction (entanglement) to individual higher level (mixed) realisations and extension (disentanglement) to the quasi-free lower-level entities/realisations: the interaction is really 'strong' in the phase of system reduction to one of its compound realisations (forming a physical 'point' of a generalised trajectory), but changes for the transient 'asymptotic freedom' in the indispensable transitional phase of $\psi$-function. Therefore any generic, 'nonseparable' interaction is *always* a *strong* interaction, irrespective of its 'average' magnitude, and at the same time cannot escape passing periodically by a phase of *separated* interacting entities (whereas, in a pathological 'separable' system this latter state is permanent and therefore the real, profound interaction is actually absent). It is evident that $\psi$-function describes simultaneously the jumps between realisations and the distribution of probabilities of individual realisations, which provides substantiation and generalisation of quantum-mechanical Born's probability rule and the practical way to find the probabilities by solving the generalised Schroedinger equation (the latter replaces all its linear-science imitations in the form of various equations for the 'distribution functions', 'density matrix', etc.). Without the composite, and permanently dynamically 'rebuilt', structure of reality made of elements-realisations which are 'gently' held together by the 'adhesive action' of permanent chaotic transitions between them, the construction of the universe would be neither strong, nor flexible enough to be viable, similar to the principle of natural and artificial 'composite materials'. Any 'self-organisation' and real 'structure formation' are impossible without this irreducibly probabilistic, multilevel complex-dynamical composite of randomness and regularity inside them.

Note also that the self-sustained, hierarchically structured, intrinsically probabilistic and dynamically entangled development of complexity of a system can be considered as the causally complete description of the generalised *adaptability* of (multilevel) complex dynamics realising the automatically maintained search by the system of the 'easiest' way of development which corresponds to the locally maximal rate of complexity unfolding and provides the latter with the fundamental property of *sense*. The generalised adaptability is closely related to the generalised 'second law of thermodynamics' and in particular its formulation for the 'freely expanding gas' which should always take the 'whole accessible volume'. In other terms, the same group of properties can be called 'generalised dynamical percolation' of the unfolding complexity, with the evident underlying interpretation involving the dynamical fractal concept.

This qualitative picture of complexity development and its formal description by the two dual versions of the Lagrange-Hamilton equation seem to be absolutely universal. Especially remarkable is the fact that the whole unified description of the entire diversity of the world, including the extended, causally complete versions of all correct notions, approaches and equations together with the ensuing results, is based on a *single* formal limitation of generality (complexity conservation law) which is shown, however, to be equivalent to all the main conservation laws and other basic principles and therefore *universally* confirmed by the *whole* body of the existing empirical knowledge (in addition to the internal consistency argument, also necessitating the existence of such kind of general order in a holistic, developing system). This is what indeed can be called the first-principle Theory of Everything (it should automatically provide the ultimate unification of knowledge, i. e. its intrinsically continuous and complete version, see also below).

Finally, the universal concept of complexity and its generalised description are applied to a brief outline of the higher levels of complex behaviour, from complex physical systems to the highest spheres of human activity, usually studied within the humanities.

It is shown that a number of most important and urgent problems, actually stagnating because of the apparently 'irresolvable' difficulties, can be solved only within the science of complexity and therefore indeed cannot yield to any canonical, unitary approach, whatever is the quantity of the applied efforts. This is precisely the objective sense of the 'end of science' (Horgan 1996): it is the end of the *linear* (dynamically single-valued, unitary) science, so that there is no any positive sense to continue the infinite fruitless 'studies' of the fundamental problems within the conventional single-valued paradigm of the scholar science which are inevitably reduced to the characteristic 'imitative' activity with fictitious, obviously contradictory results formally 'certified' by purely subjective, 'organisational' methods of voting, traffic of influence, etc. In contrast, the complex-dynamical solutions are universally feasible, objectively verifiable and complete by their origin, which means also that each of them contains a much larger scope of qualitatively different possibilities with respect to everything that could be expected within the linear approach. This conclusion is illustrated by such problems and directions as: the many-body problem and modern solid state theory that includes the well-known evidently stagnating difficulties like the origin of the high-temperature superconductivity, systems with frustration (glasses), quantum Hall effect and other situations with irreducibly strong, nonperturbational interaction effects; 'megaprojects' of the canonical science, including 'supercomputers', 'superaccelerators', new energy sources (especially the controlled nuclear fusion) and the human genome project; quantum computers, neural networks, artificial intelligence, and the problem of consciousness; medical, ecological and geophysical problems (transition to objectively efficient integral medicine and ecological system monitoring, efficient prediction and control of natural catastrophes are outlined); causally complete understanding of economical, sociological, and political development of the modern society; new organisation of science in relation to the new contents and meaning of the intellectual activity at the level of the universal science of complexity; rigorous interpretation of aesthetic and ethical notions, theological concepts, and the role of philosophy. Using the power of the qualitatively new paradigm of dynamic redundance, the universal science of complexity provides the genuine, causally complete, first-principle *understanding* of the natural processes at any level of complexity forming the unique basis for their efficient practical monitoring, whereas the canonical single-valued (i. e. essentially one-dimensional) analysis inevitably limits itself to the purely empirical method of positivistic classification of observations with the help of unexplained postulates, which excludes any possibility of *causally* substantiated and therefore *objectively* reliable predictions. It is not surprising that this essential difference comes in the foreground and becomes vitally important just today, when the civilisation development brings the centre of human activity to superior levels of complexity: now we *need* to *master* any dynamic behaviour, which is *equivalent* to its complete, realistic, totally conscious *understanding* uniquely based on the *extended causality* of the universal science of complexity.

The possibility or impossibility of reduction, at least partial, of the full complexity of a system to a single-valued, regular behaviour determines its classification (intuitive until now) as the subject of respectively 'exact sciences' or the humanities (arts). It is natural that the latter type of knowledge seems to be basically incomprehensible at the completely conscious, properly 'scientific' level within the canonical paradigm, and belongs therefore to the irreducibly empirical, intuitive 'science', or *art*, even though all the conventional 'exact' sciences also inevitably contain excessive quantities of basically incomprehensible, postulated elements (the 'mysteries' of quantum mechanics provide a characteristic example). It is equally clear why the exact sciences typically study the lower levels of the universal hierarchy of complexity of the world, whereas the humanities tend to its higher levels (with the 'natural sciences' forming the irregular border between the two).

The existing mathematics totally belongs to the single-valued paradigm of the canonical science and therefore *should* be *irreducibly* detached from the intrinsically multivalued reality in order to attenuate the evident divergence with it, as it is so transparently illustrated by comparison of the cumbersome, rigidly fixed, simplified constructions of the canonical 'maths' with the infinitely fine involvement of the always *moving* (self-developing), unpredictable and asymmetric structure of the world around us. For the same fundamental reason the canonical mathematics cannot be universal and noncontradictory in its internal structure, but should inevitably be composed from the disrupted, often *antagonistically* opposed notions and structures. By contrast to this, the *new mathematics* of the universal science of complexity is intrinsically *entangled with reality* at all scales, which means that now all the mathematical constructions are built only in close *direct* relation to well-specified real structures. In fact, the new mathematics always deals with one and the same construction, indistinguishable from the reality it describes - the ever developing fundamental dynamical fractal of the world - possessing infinitely many aspects and branches which now are directly related and opposed to each other only *constructively* (dualistically), and not antagonistically. Therefore in the new mathematics one always has a well-defined, though nontrivially structured, 'guiding line' for development of formal description which assures automatically the maximum efficiency, since it coincides with the universal hierarchy of complexity *including* its own formal representations. In practice, one should simply follow the natural complexity unfolding process using the unique criterion of (dynamic) complexity conservation which includes permanent growth of its structural part (generalised entropy) and the intrinsic continuity, typically appearing through a system of truly noncontradictory, multi-sided and multi-level, *correlations* within the unique hierarchy of complexity (the natural process of image recognition by the brain provides a pertinent analogy). One can compare this new type of formal description with the usual method of intuitive, often purely abstract 'guesses' about the 'suitable' development of formal structures followed by their 'experimental verification' in the applied canonical mathematics or 'internal consistency' checking in the pure mathematics which lose so definitely and hopelessly their relation to reality and the ensuing 'certainty' about the objective criteria of truth (cf. Feynman (1965), Kline (1980), Ziman (1996)), not accidentally just in the age of the crucial advent of the irreducible dynamic complexity of the higher levels of being.

This new paradigm of dynamic cognition actually refers to the whole process of *new knowledge development*, which means that in the universal science of complexity there can be no any discontinuous gap between 'exact', or 'formal', and 'figurative' (verbal) descriptions; they are intrinsically unified within the single, ultimately causal, really complete (and still ever developing!) *understanding*. *Knowledge* itself is causally defined now as a part of the universal hierarchy of tangible complexity (*entropy*) concentrated rather at its higher levels and emerging in the same universal process of natural transformation of dynamic information (latent interaction) into entropy. It is clearly demonstrated in this work that the universal science of complexity provides a completely new type of cognitive process leading to the *ultimate unification of knowledge*, which means that the obtained knowledge is intrinsically, indivisibly unified, in perfect accord with the unity of nature it describes. The *structure* of the unified knowledge comes not from the artificial, mechanistic 'classification' into basically separated 'fields' and entities, as it happens in the canonical science, but from the naturally appearing levels of complexity of the actually observed objects and dynamical patterns of their behaviour. The naturally emerging, 'living' continuity of the new knowledge endows the related new thinking with the fundamentally important property of the *ultimate, and real, freedom* and intrinsic *globality* understood as a really free, individually specified 'walk' *within* the *whole* multilevel arborescence of complexity which remains at the same time always *meaningful* (objectively correct) due to the universal *criterion of truth* (the growing dynamic complexity within the hierarchy of knowledge). Similar to the reality it describes, the universal and living knowledge of the science of complexity has the *dynamically entangled*, fractal-like, and partially unpredictable, always *developing* structure. This latter property emphasises the *intrinsically creative* character of the new knowledge, as opposed to the basically fixed, chronically stagnating character of the canonical science always actively *resisting* to *any* qualitative change (cf. Kuhn (1970)). The realisation-concepts of the unreduced, complex-dynamical knowledge-entropy, including their *new* mathematical representation in the form of (extended) *equations*, interact (entangle) with one another, *giving birth* to dynamically uncertain and therefore chaotically searching, intrinsically 'asymmetric' notions (equations) of the emerging higher level(s), in *full* correspondence with respective stages of the universal process of complexity development in nature. The intrinsic *completeness* of this new, living science does not contradict to its permanently *changing* structure: the self-developing knowledge-entropy always 'tries' to 'percolate' into the largest accessible 'space' under the 'pressure' of the latent interactions (dynamic information), so that the *relative* completeness of the already attained levels of complexity, including their complete (ultimate) unification, forms the necessary *basis* for the *development* of the next levels. The ultimate unification of knowledge within the science of complexity is therefore both a particular state and the general process of change providing the most complete description of the world, which actually cannot, and should not, be separated from the developing reality it describes, both are equally 'objective' and creative, and form the unbroken, *unique Truth*. This intrinsic realism of the unreduced science of complexity is another expression of its *extended causality* which includes natural reunification of the 'three worlds' (physical, mental, and mathematical) artificially separated within the canonical linear thinking (cf. Penrose (1994)).

It is not surprising that this highest form of knowledge is capable to provide the objective, causal understanding even for such parts of reality, 'esoteric' within the canonical science, as *consciousness*, *ethical* and *aesthetic* notions it produces, or a basically different, 'supernatural' reality usually studied by *theology*.

In particular, it is rigorously shown that the universal complexity conservation law, supported by all existing observations, directly leads to the necessity of explicit *Creation of the World* from the *outside*, represented by more fundamental levels of dynamic complexity, in the form of the primal protofields and their interaction, as opposed to the linear-science concepts of 'spontaneous' emergence of the world 'from nothing', formally permitted by the canonical energy conservation law and serving as a basis for various *'ex nihilo'* concepts dominating in the linear-science cosmology (see e. g. Gott and Li (1997) and the references therein). This conclusion is related to the fact that the unreduced dynamic complexity-energy is always *positive* (and large, for the whole World), contrary to the mechanistically interpreted energy from the linear science. On the other hand, the dynamic redundance paradigm of the new science of complexity explains the source and content of the autonomous internal development (unfolding) of the once created complexity of the world revealing its *raison d'etre* and the objective goal (sense) and content of its evolution. The existing painful separation between the main directions of progressive theological thought, such as those of Christian, Islamic, and Buddhist traditions, can be replaced, within the extended causality of the universal science of complexity, by a *constructive* interaction leading to crucial progress in each of these branches partially demonstrated in this work and avoiding any mechanistic mixing of the linear-thinking substitutes sometimes actively promoted today.

The property of *consciousness* emerges together with certain sufficiently high levels of the unfolding complexity (provided the initially created quantity of its informational form is high enough). Similar to any other form of complexity, it is neither strictly detached from 'unconscious' cognition at the lower border, nor limited from above; it contains many qualitatively different, inhomogeneously distributed levels and 'branches'. The 'non-computability of conscious thinking' (Penrose 1994) cannot be considered as a specific attribute of consciousness simply because *any* truly complex (and thus practically *any real*) behaviour, starting from the isolated electron dynamics, involves *causal randomness* and is therefore beyond any linear-science 'computability'; consciousness is a *higher-level* 'non-computable' behaviour. Its human version, the only one currently known to us, is physically realised as higher levels of the complex dynamics of the brain which can be described as a hierarchy of objective reduction-extension processes of the realistic 'brainfunction' representing the universal $\psi$-function at this level of complexity and describing the distribution of the magnitude of the physically real, basically well-known electro-chemical interactions between the neurones. At the corresponding (higher) levels of brain dynamics, those chaotic reductions take the form of partially unpredictable appearance of 'feelings', 'thoughts', or 'ideas' that 'suddenly cross one's mind', which provides also a transparent manifestation of the property of *creativity* inherent to any unreduced complex dynamics (thus a 'new idea' results from a *physically real* process of chaotic entanglement-disentanglement pulsation between the interacting ideas/realisations from a preceding lower level of thinking). The mysteriously detached from the physical world, postulated *res cogitans* is now causally explained as a *res extensa* from a high enough level of the *universal* complexity. In accord with universality of complex behaviour within the dynamic redundance paradigm, the brainfunction dynamics generally resembles the dynamics of the wavefunction in (the extended) quantum mechanics, but this is only a behavioural, though quite significative, analogy: there is *no* any *direct* involvement of the lowest levels of complexity, described as quantum (wave) mechanics, in the highest levels of complexity associated with brain functioning (cf. Hameroff and Penrose (1995)), even though the fact that we can describe both ultimately separated levels of behaviour within the same general concept, as well as the underlying profound analogy between them, are quite remarkable and convincingly demonstrate once more the inimitable universality of the proposed concept of complexity.

The property of being *beautiful*, attributed to a complex system, can be consistently *quantified* within the universal science of complexity by a value proportional to its *actually perceived* dynamic complexity-entropy expressed by the number of actually observed realisations (rather than the rate of their change). Correspondingly, the beauty of an *image*, and in particular of a static one which formally has zero dynamic complexity in itself, is determined by the effective dynamic complexity of the process resulting from interaction between the image and an *observer* and existing within the observer that should be represented therefore by at least a multilevel, nonequilibrium (i. e. alive, see above), but *not necessarily* conscious, complex system. This definition is consistent with the well-known *subjectivity* of esthetical estimates which thus can also be rigorously explained, quantified, and reduced to an *objectively* determined minimum, if necessary.

Similarly, the ethical (and theological) notions of *good and evil* can be provided with a universally applied, objective interpretation and the ensuing rigorous quantification, without any simplification of their intrinsic involvement. *Good* is proportional to complexity-entropy *growth* in the process of its natural emergence from the latent, informational form described above. Correspondingly, *evil* should be understood as the absence (or relative slowing down) of progressive complexity unfolding, and therefore evil is not an independent property, but just the absence of good (which can only be temporal), which should be expected. Good is naturally associated with *progress* (objectively defined complexity growth), while evil is directly related to *stagnation* of development. At higher levels of complexity corresponding to human behaviour, evil often appears in the form of patterns of lower dynamic complexity *artificially* ('deliberately') *imposed* to, or substituted for, the forthcoming higher-complexity behaviour in order to suppress the 'normal' advance of good and create or prolong the opposed state of stagnation, which is profitable for the intrinsic adherents of the lower-complexity behaviour. Needless to say, any apparent 'victory' of the adherents of evil can only be ephemeral, since the pressure of the hidden stock of complexity will then destroy their 'achievements' with multiply increased, accumulated force, or else the system will die *as a whole*, if it does not contain the necessary quantity of disposable latent complexity. This observation seems to be an empirically correct illustration of the rigorously substantiated complex-dynamical interpretation of the 'eternal struggle between good and evil', between the forces of creation (genuine progress) and nothingness (stagnation, degradation, false progress and "false prophets").

The intrinsic self-consistency of the universal concept of dynamic complexity extends its validity even beyond the borders of the now perceptible, material world (not to mention the canonical forms of knowledge), and the symmetry of complexity characterises the truly universal, and therefore unique, Principle of Being, reproducing its full diversity, with all the known, imaginable, and undreamed-of richness of any intricacy and profundity (it is compared to the similarly universal concepts of the Oriental philosophy and theology, such as *Tao* or *Chu*, and used to obtain their objective refinement).

This ultimately unified picture of the universal science of complexity had a number of great precursors, even though they almost always remained in the shadow of the dominating mechanistic approach of the conventional linear thinking and suffered dramatically from its aberrations. The methods, attitudes and results of such great representatives of the *essentially* nonlinear (non-unitary) thinking as René Descartes, Henri Bergson and Louis de Broglie cannot be omitted from the history of the new, basically multivalued knowledge of the science of complexity which is thus not absolutely new and can be related, in a well-specified fashion *avoiding any mechanistic substitutions*, to other prominent names of science. This will constitute the subject of the new, objectively re-interpreted and considerably extended history of knowledge. In particular, the Cartesian science is 'rediscovered' and 'post-substantiated' now, in many its detailed aspects, as the unambiguous beginning of the unreduced and universal science of complexity which was not actually recognised as such and has later been replaced by the more and more simplified and dominating mechanistic approach, with the evident great losses resulting, in particular, in the degradation of the 'criteria of truth' (especially rapid in course of the twentieth century), up to the current fundamental impasse of the 'end of science'. The canonical mechanistic approach has perfidiously endowed the extended Cartesian thinking with its own, 'one-dimensional' limitations, whereas the unreduced essence of the science of complexity, clearly emphasised by René Descartes, is based on the idea about intrinsically chaotic (dynamically redundant), self-interacting, and therefore autonomously developing 'mechanism' of the world and any its part.

The new knowledge of the universal science of complexity is definitely not the positivistic 'arrangement of impressions', a dull stock of immobile, rigidly fixed facts, constantly enriched by new ones. It is a *living*, permanently and *naturally* changing *organism*, realising *another way of thinking* accessible and destined to *everyone* and not only to a narrow group of self-'chosen', well-organised 'sages' hiding their mediocrity and selfish desire to dominate behind incomprehensible systems of technical sophistication and oligarchic organisational structure, as it happens today in the canonical, unitary science. Therefore the new knowledge of the universal science of complexity is inseparable from the new, objectively founded *criterion of truth* consisting in emergence of the causally complete (and therefore always dynamically probabilistic), *real solutions for real problems* characterised by a *holistic system of correlations* of the unreduced and always *growing* dynamic complexity.

Being applied to the currently acute problem of *organisation of science* this new criterion and approach of the unreduced science of complexity show unambiguously that the existing organisational structures of the totalitarian, self-estimating type have entered in a fundamental, antagonistic contradiction with the demand of progressive, truly creative development of any form of knowledge, but especially that of the universal science of complexity, not accidentally stepping out to the foreground of any creative search just in the epoch of the 'end' of the canonical, mechanistic science and the corresponding unitary type of its organisation. Moreover, the same objectively based approach shows a way and provides the unique *fundamental* substantiation for the new, qualitatively different type of organisational structure really *adequately* corresponding to the forthcoming stage of development at the necessarily much *higher* level of complexity. The new organisation of science - and eventually of the *whole society* - can only be based on an explicitly *non-unitary*, really freely developing, informatively open and accessible to every one, *distributed system* of dynamically interacting, but *independent*, typically small 'enterprises' or 'units'. The *openness* means in particular that the success of activity of each of them and of any their group will be dynamically estimated by many interconnected, *actually* independent partners looking for the *actual emergence* of *causally complete* solutions to *real* problems (practical or fundamental), the solutions and their estimations being *explicitly presented* and easily accessible to everybody (e. g. via electronic networks). The basically rigid, centralised subordination of the unitary organisational structure, including any its self-preserving 'progressive' variations which change nothing in the base, is replaced here with the distributed *individual responsibility* and *motivation* of *each one* of the participants, without eliminating the *autonomously* controlled, *naturally* evolving dynamical structure of the system.

As it becomes especially obvious within the universally substantiated analysis of dynamic complexity, the mechanistic scholar science is incapable to resolve any nontrivial problem for the *same* fundamental reasons as those that do not permit the 'one-dimensional' rationalism of the Unitary Thinking to realise the full potential of human consciousness in a truly *developed* form of civilisation providing *everyone* with a *realistic* possibility of (almost) *complete* realisation of his *individually* specific potential complexity. The purely technical (mechanistic) way of development, unavoidable within the dominating Unitary System of organisation, cannot lead to resolution of the urgent fundamental problems either in science, or in life in general, however large are the quantities of the applied efforts (this is the basically inevitable *saturated incompleteness* of a fully developed level of complexity). Correspondingly, the new level of thinking of the universal science of complexity inevitably involves the crucial progress towards a qualitatively superior level of consciousness and thus opens the unique way to another level of life (complexity), where one shall be able to naturally solve the complicated, 'irresolvable' practical problems accumulating now in all spheres of life. It is the *only* possible and *urgently* needed way of further *progressive* development, without which the civilisation is seriously risking to be destroyed as it is clear from the objective complex-dynamical interpretation of the easily recognisable current exhaustion of the 'developed' world (the End), forming the local (bifurcational), but extremely important, Apocalyptic end of a very big stage of complexity unfolding.

This fundamentally stagnating, ultimately decadent character of the modern epoch of the End is clearly 'felt', in various ways, at the empirical level. The essential contribution of the unreduced analysis of the science of complexity is that it permits one to understand and objectively substantiate the ultimate origin of development and the ensuing reason for its current stagnation. If the development is determined by the unfolding system complexity, then the current stagnation of the process is due precisely to recent crucial, fundamental *success* of especially its *technical* branches permitting, *for the first time in history*, to provide *everybody with everything* necessary to satisfy the maximum of objectively normal material human demands and even much more than this. However, already the actual completion of this stage of development (to say nothing about the ascent to its higher stages) is impossible within the *same* level of complexity, since it would actually produce its absolute saturation (due to absolute personal satisfaction and the ensuing absence of demand), which is equivalent to the dynamical death of the system. The currently used means to resolve a problem within the same level of complexity inevitably involve various forms of *artificial* limitation of the level of life for large enough layers of population, but this can only increase the deepness of the impasse of development manifesting itself as the *permanent universal crisis* of every aspect of life, including many irregular variations due to the growing intrinsic *instability* of the stagnating system. The basically complete analysis of the universal science of complexity shows that the only right solution leading to the issue from the impasse should involve the explicitly performed, profound and many-sided transition to the next level of complexity oriented towards satisfaction of higher, non-material needs of each personality and thus being practically unlimited 'from above'. This will demand a serious displacement of accents in the existing distribution of efforts towards the *intrinsically complete* individual *self-realisation* for *every one* which can only be based on the omnipresent, easily accessible, unreduced *creativity* and should include the transition to the new, non-unitary type of social structure and the related new way of thinking outlined above. Correspondingly, this *truly* developed society should definitely abandon such 'eternal' properties of the 'developed' unitary way of life as materially based formal ambitions motivating the antagonistic struggle by any accessible means for better 'places' and 'things', deceitfully hidden behind the officially proclaimed, purely formal 'liberty', 'equality', and 'fraternity'. In particular, the bureaucratic, mechanistic, and self-seeking way of various 'unifications' within Europe and the World currently promoted by the governing oligarchies actually leads only to further deepening of the existing antagonistic ruptures between the 'subjects' of unification, nations and individuals, and should be replaced by the *direct*, *creativity*-driven interaction of those subjects estimated by, and *only* by, the emerging *results* of creation. Such are the most essential features of the forthcoming unavoidable 'boost' of complexity development beginning with the Revolution of Complexity that initiates the transition from unitary to explicitly complex-dynamical, 'non-computable' way of conscious thinking.

# Epilogue: *The Beginning*

Complexity, nonlinearity, chaos, self-organisation, criticality... The frustration of the incomprehensible, disordered agglomerate of contradictory ideas turns now into the majestic harmony of the Distributed Creation. The Universal Science of Complexity provides each notion with a well-specified, profound sense and simultaneously unifies all of them in the holistic, self-consistent structure of reality, possessing the unique capacity to transform the intrinsic conflicts into constructive interactions which use the energy of formal opposition as the creative power of the unfolding Potential of Complexity. The tamed Dualities between the main Entities, which are the forms of the dynamic complexity, and their Properties, characterising the reality of their emergence, explode in the multiplicity of hierarchically organised structures which constitute the natural richness of Being.

This *Ultimate Unification of Knowledge* is the Beginning of the new science, inseparable from the new history and new level of life, in all its manifestations. The completely unified Truth cannot belong to several separated worlds or realities, it is created by the fine and always growing, probabilistic entanglement of the 'heavy' material of 'tangible' reality with the fine tissue of causal knowledge and searching imagination.

Every part and the total contents of the Truth can be expressed by the universal and rigorous mathematical formalism involving, however, quite New Mathematics. Or if one prefers, the same Truth can be expressed by other, 'figurative' means using the power of the New Arts, and this 'non-formal', 'artistic' description is generally not less rigorous. It is so because New Mathematics and New Art are not opposed any more, as neither are knowledge and reality, they are simply different, but explicitly related, aspects of the same integral construction of Truth, each possessing its own, local completeness and particular, well-specified features. This is the *holographic* Truth-Reality, where each part or aspect reproduces the whole, but their totality provides the 'best quality of the image', the ultimate completeness and Causality.

The emerging New Reality has nothing to do with the hypocritical imitation of justice and progress of the End, either in the scholar, linear science, or in the basically stagnating, completely exhausted Unitary System of social organisation in all its 'democratic', 'meritocratic', and 'totalitarian' versions always degenerating into a grey, passive 'equality' which is actually reduced to a low-level mediocrity and hides the ugly, frustrating antagonism. The Beginning leads to the genuine Creativity that can only originate from the properly specified, constructive Interaction between the suitably *differing* partners and tendencies naturally 'chosen' by and *only* by their *intrinsic, real capacities* (which are *objectively* defined by the stock of the latent complexity and subsequently confirmed by the *actually* realised *progress*, the unambiguous *increase* of the explicit, unreduced complexity of the *realistic* new possibilities). The exhausted, formal liberté-égalité-fraternité, resulting in the actual absence of the *real* free choice and thoroughly maintained by the mechanistic power of the Technocracy of the End motivated by the selfish profits it takes from the System, is replaced by *Volonté* (for the Choice), *Créativité* (for the Realisation), *Diversité* (for the Result) of the Beginning.

The Beginning needs the End, as its irreducible dual opponent giving sense to its own emergence. They are both conceived from the start, they have always been existing, fighting, disappearing and reappearing at the changeable interface between the Past and the Future. The Medieval Thinking of the End always uses the inertial power of mediocrity to maintain the mechanistic dominance of the Past over the creative force of Renaissance Thinking advancing the World to the Beginning Future.

The Beginning starts always, it is in the advent of prophets and any other Breakthrough of the same Universal Truth culminating in the Values of Renaissance and emerging in each partial advance of progress, whenever and wherever it happens. The Values prepare the decisive, global Beginning that needs Liberation. Liberation is the meaning of the End.

The Beginning is a victory of Good that corresponds to the crucial *growth* of complexity and realises another big step in progressive unfolding of Creation. But creativity needs independent, autonomous interactivity, and therefore the Beginning results from the End in the fight between Good and Evil. Evil is the Inertia of the Past, a lower-level complexity, a 'yesterday triumph' that does not want to go easily to the back-yard of Being, often masking itself behind artificially simulated substitutions for 'progress', 'interaction', 'unification' which are totally composed from the superficial, abstract signs and in reality lead only to a more profound stagnation: "Beware of false prophets". Nothingness is their choice, emptiness is their way, oblivion will be their fate.

The Change for Good is what has ever been happening in course of the eternal complexity development, providing the sense for Time as the unique form of the World existence. But the change can never emerge smoothly, predictably and with the efforts of 'someone else', though many yield to the deceptive comfort of this illusion. The End shows that the illusions are vain, but only the Beginning can kill them by actually creating another reality with much higher complexity that just expels the Evil to Yesterday, with all its tricky substitutes, hypocritical 'tolerance' and perfidious 'good manners'.

The Beginning prepared by Liberation needs the definite, active Choice leading to crucial Change for Good, here and now. It is impossible without the decisive restriction of the interests of Evil, which implies also clear rejection of the imitations of progress it uses, and the sooner it is performed, the less dramatic will be the consequences: the 'bridges to the third millennium' hastily pumped up by the servants of Evil at their bazaar level of complexity lead to a deep impasse and the ensuing catastrophe. The Alternatives of the Choice, representing today's huge Bifurcation of Development, are *within each personality* bearing a germ of the Renaissance Thinking, and by making the right choice every one leaves the End of a clever animal and ascends to the Beginning of his true, unique Self, lit up by the incarnate spirituality.

Similar to some fantastic Star Gate, the real Portal between the Past and the Future is wide open only during a limited period of Time, simply because Time is inhomogeneous and irreversible. Something will necessarily happen, and if it is not the Good of the Beginning that needs to be actively promoted, it will be its absence, the Evil of the supersaturated End, which gives the catastrophic way of development whatever is the apparent 'technical progress'.

There is indeed a time to every thing. The time of lie slips irreversibly to the past, and now it is the time of truth waiting for *action*. One cannot make the necessary big advance towards the Truth remaining in conformity with the lie of the Unitary System of life. The banality of the most sophisticated lies becomes explicitly evident through the direct comparison with the intrinsically humanised, irrestrictably interesting, always dualistic and developing truth of the ultimately unified knowledge.

The Beginning is here and the objective, fundamentally substantiated analysis of the Universal Science of Complexity makes it clearly visible from the End through the opening Portal to the Future. You are at the very Threshold of a quite New World prepared by the whole now ending History and especially by the last period of Anno Domini. You just need to make a step, You who are reading now these lines, and not someone else, supposed to be less occupied with the vain flicker of the End. You will know for whom the Bell tolls, You just should not reject its Tune. The Cynicism of the condemned System will not protect You, it leads only to nothingness.

The Step gives the true causality and complexity in science and another level of life, oriented to the real and complete realisation of your possibilities which are not opposed any more to those of the others, as it is artificially imposed by the Evil of the dying System.

It is the decisive step towards the true, creative interactivity and entanglement between fields of knowledge, nations and people that should replace the mechanistic, superficial substitutions for unity hastily prompted right now by the decadent, mercenary Evil that belongs to the Past already for at least two thousand years.

That is *the only* one *realistic* way of *progressive* development opposed to the deceptive easiness of multiple modes of passive degradation à la 'let it be'. The Step needs the efforts because it is an Act of Creation, a victory of Good over Evil. But if everything is prepared for it and all the tendencies point in the same direction, there is no sense to refuse from one's own Future, especially when it is so sensitive to one's action and can be so Good.

\*\*\*\*\*\*\*\*\*\*\*\*\*\*\*\*\*\*\*\*\*\*\*\*\*\*\*\*\*\*\*\*\*\*\*\*\*\*\*\*\*\*\*\*\*\*\*\*

*Ask,*
*and it shall be given you;*
*Seek,*
*and ye shall find;*
*Knock,*
*and it shall be opened unto you:*

*For every one that asketh receiveth;*

*And he that seeketh findeth;*

*And to him that knocketh it shall be opened.*

Matthew 7:7-8

\*\*\*\*\*\*\*\*\*\*\*\*\*\*\*\*\*\*\*\*\*\*\*\*\*\*\*\*\*\*\*\*\*\*\*\*\*\*\*\*\*\*\*\*\*\*\*\*

# Appendix

TABLE OF CONTENTS OF THE BOOK

## Part II: Complexity in Arbitrary Dynamical Systems: The Unified Description

# Conclusion